\documentclass[prb,showpacs,preprint,nofootinbib,nofloats,endfloats]{revtex4}

\usepackage{amsmath,amssymb,epsfig,epsf}

\begin{document}

\title
{\bf Pseudogaps in Strongly Correlated Metals: Optical Conductivity within the
Generalized Dynamical Mean-Field Theory Approach}
\author{E.Z. Kuchinskii, I.A. Nekrasov, M.V. Sadovskii}

\affiliation
{Institute for Electrophysics, Russian Academy of Sciences,
Ekaterinburg, 620016, Russia}

\begin{abstract}
Optical conductivity of the weakly doped two-dimensional repulsive 
Hubbard model on the square lattice with nearest and next nearest hoppings
is calculated within the generalized dynamical--mean field 
(DMFT+$\Sigma_{\bf p}$) approach which includes correlation length scale $\xi$
into the standard DMFT equations via the momentum dependent 
self--energy $\Sigma_{\bf p}$, with full account of appropriate vertex 
corrections.  
This approach takes into consideration non-local dynamical 
correlations induced e.g. by short--ranged collective SDW--like 
antiferromagnetic spin fluctuations, which (at high 
enough temperatures) can be viewed as a quenched Gaussian random field with 
finite correlation length $\xi$. 
The DMFT effective single impurity problem is solved by 
numerical renormalization group (NRG).  We consider both the case of 
correlated metal with the bandwidth $W\lesssim U$ and that of doped Mott 
insulator with $U\gg W$ ($U$ --- value of local Hubbard interaction). Optical 
conductivity calculated within DMFT+$\Sigma_{\bf p}$ demonstrates 
typical pseudogap behavior within the quasiparticle band in 
qualitative agreement with experiments in copper oxide superconductors.
For large values of $U$ pseudogap anomalies are effectively suppressed.  
\end{abstract}

\pacs{71.10.Fd, 71.10.Hf, 71.27+a, 71.30.+h, 74.72.-h}

\maketitle

\newpage

\section{Introduction}

Pseudogap state is a major anomaly of electronic properties of
of underdoped copper oxides \cite{Tim,MS}. We believe that the preferable 
``scenario'' for its formation is most likely based on the model of strong 
scattering of electrons by short--ranged antiferromagnetic (AFM, SDW) 
spin fluctuations \cite{MS}. This scattering mainly
transfers momenta of the order of ${\bf Q}=(\frac{\pi}{a},\frac{\pi}{a})$ 
($a$ --- lattice constant of two dimensional lattice) leading 
to the formation of structures in the one-particle spectrum, 
which are precursors of the changes in the spectra due
to long--range AFM order (period doubling) with
non--Fermi liquid like behavior 
of spectral density in the vicinity of the so called ``hot-spots'' on the
Fermi surface, appearing at intersections of the Fermi surface 
with antiferromagnetic Brillouin zone boundary (umklapp surface) \cite{MS}.

In recent years a simplified model of the pseudogap 
state was studied \cite{MS,Sch,KS} under the assumption that the scattering
by dynamic spin fluctuations can be reduced for high enough temperatures
to a static Gaussian random field (quenched disorder) of pseudogap fluctuations.
These fluctuations are defined by a characteristic scattering vectors of the 
order of ${\bf Q}$,  with distribution width   
determined by the inverse correlation length of short--range
order $\kappa=\xi^{-1}$,  and by appropriate energy
scale $\Delta$ (typically of the order of crossover temperature $T^*$ to 
the pseudogap state \cite{MS}). 

It is also well known that undoped cuprates are antiferromagnetic Mott 
insulators with $U\gg W$ ($U$ --- value of local Hubbard interaction, $W$ --- 
bandwidth of non--interacting band), so that correlation effects are very 
important and underdoped (and probably also optimally doped) cuprates are 
actually typical strongly correlated metals. 

The cornerstone of the modern theory of strongly correlated systems is 
the dynamical mean--field theory (DMFT) 
\cite{MetzVoll89,vollha93,pruschke,georges96,PT}. At the same time, standard
DMFT is not appropriate for the
``antiferromagnetic'' scenario of pseudogap formation in strongly
correlated metals due to the basic approximation of the DMFT, which
completely neglects non-local dynamical correlation effects.

Different extensions of DMFT were proposed in recent years to cure this
deficiency, such as extended DMFT (EDMFT)\cite{Si96,Haule}, which locally includes coupling
to non-local dynamical fluctuations, and, most importantly, different versions
the so-called cluster mean-field theories, such as 
the dynamical cluster approximation (DCA)\cite{TMrmp} and
cellular DMFT (CDMFT)\cite{KSPB}. 
However, these approaches have certain drawbacks.  
First of all, the effective quantum single
impurity problem becomes rather complex. Thus, majority of computational
tools available for the DMFT can be used only for small enough clusters
\cite{TMrmp}, i.e.\ include mostly nearest-neighbor fluctuations. It is
especially difficult to apply these methods to calculations of 
two--particle properties, e.g. optical conductivity.

Recently we have proposed generalized DMFT+$\Sigma_{\bf p}$
approach \cite{JTL05,PRB05,FNT06}, which
on the one hand retains the single-impurity description of the DMFT, 
with a proper account for {\em local} correlations and the possibility 
to use impurity solvers like NRG\cite{NRG,BPH}, while on the 
other hand, includes non-local correlations on a non-perturbative model 
basis, which allows to control characteristic scales and also types of 
non-local fluctuations. This latter point allows for a systematical study of 
the influence of non-local fluctuations on the electronic properties and in 
particular provides valuable hints on physical origin and possible 
interpretation of results. Within this approach we have studied single --
particle properties, such as pseudogap
formation in the density of states of the quasiparticle band both for
correlated metal and doped Mott insulator, evolution of non--Fermi liquid like
spectral density and ARPES spectra \cite{PRB05}, ``destruction'' of
Fermi surfaces and formation of Fermi ``arcs'' \cite{JTL05}, as well as
impurity scattering effects \cite{FNT06}. This formalism was also combined 
with modern LDA+DMFT calculations of electronic structure of ``realistic''
correlated systems to formulate LDA+DMFT+$\Sigma_{\bf p}$ approach, 
which was applied for the description of pseudogap behavior in 
Bi${_2}$Ca${_2}$SrCu${_2}$O${_8}$ \cite{cm06}.

In this paper we develop our DMFT+$\Sigma_{\bf p}$ approach for 
calculations of two--particle properties, such as (dynamic) optical 
conductivity, which is conveniently calculated within the standard DMFT
\cite{pruschke,georges96}. We show that inclusion of non-local correlations
(pseudogap fluctuations) with characteristic length scale $\xi$ allows 
the description of pseudogap effects in longitudinal conductivity of the
two-dimensional Hubbard plane.

The paper is organized as follows:
In section \ref{leng_intro} we present a short description of our 
DMFT+$\Sigma_{\bf p}$ approach. In section \ref{opt_cond} we derive basic
DMFT+$\Sigma_{\bf p}$ expressions for dynamic (optical) conductivity, as well
as formulate recurrence equations to calculate the {\bf p}--dependent 
self--energy and appropriate vertex part, which take into account all the 
relevant Feynman diagrams of perturbation series over pseudogap fluctuations.
Computational details and basic results for optical conductivity
are given in section \ref{results}. We also compare our results with that of
the standard DMFT. The paper is ended with a summary 
section~\ref{concl} including a short overview of related experimental 
results.

\section{Basics of DMFT+$\Sigma_{\bf p}$ approach}
\label{leng_intro}

\begin{figure}
\includegraphics[clip=true,width=0.8\textwidth]{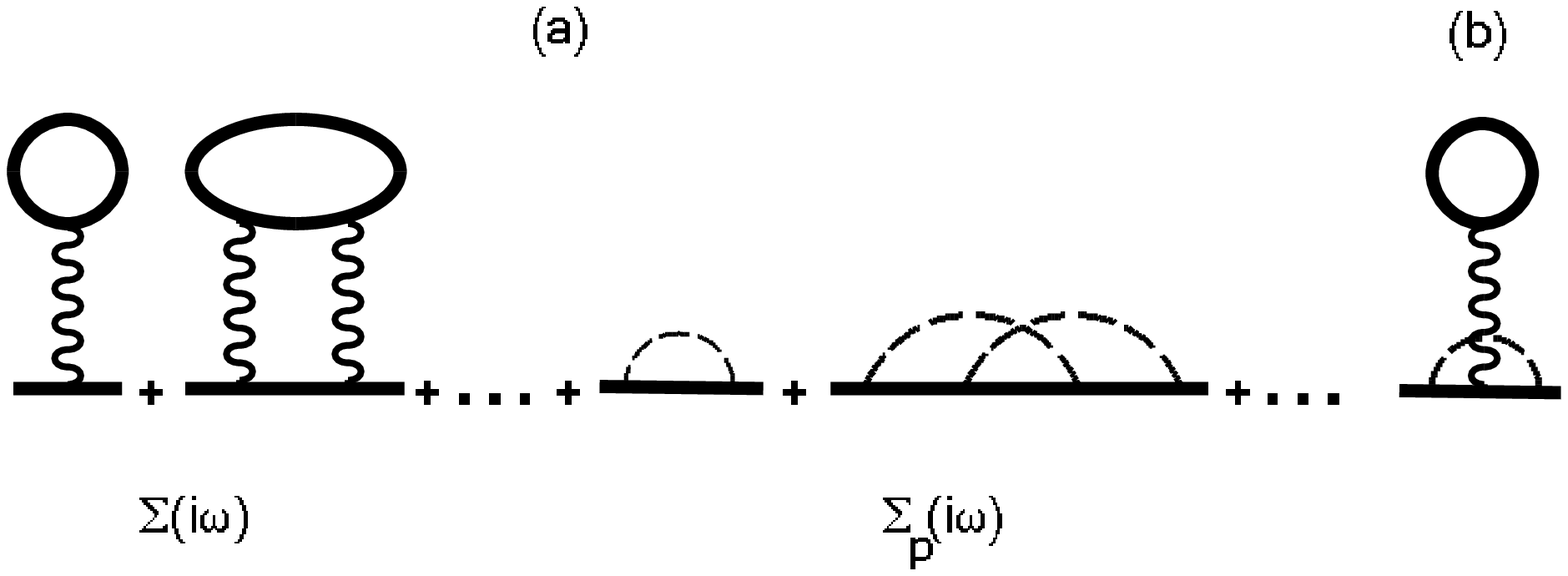}
\caption{Typical ``skeleton'' diagrams for the self--energy in the
DMFT+$\Sigma_{\bf p}$ approach.
The first two terms are examples of DMFT self--energy diagrams; the
middle two diagrams show some contributions to the
non-local part of the self--energy (e.g. from
spin fluctuations) represented as dashed lines;
the last diagram (b) is an example of neglected diagram's leading to
interference between the local and non-local parts. }
\label{dDMFT_PG}
\end{figure}

As noted above the basic shortcoming of the traditional DMFT approach
\cite{MetzVoll89,vollha93,pruschke,georges96,PT}
is the neglect of momentum dependence of electron self--energy.
To include non--local effects, while remaining within the usual ``single impurity
analogy'', we have proposed \cite{JTL05,PRB05,FNT06} the following 
(DMFT+$\Sigma_{\bf p}$) approach. First of all, 
Matsubara ``time'' Fourier transformed single-particle Green function 
of the Hubbard model in obvious notations is written as:
\begin{equation}
G(i\varepsilon,{\bf p})=\frac{1}{i\varepsilon+\mu-\varepsilon({\bf p})-
\Sigma(i\varepsilon)
-\Sigma_{\bf p}(i\varepsilon)},\qquad \varepsilon=\pi T(2n+1),
\label{Gk}
\end{equation}
where $\Sigma(i\varepsilon)$ is the {\em local} contribution to self--energy,
of DMFT type (surviving in the limit of spatial dimensionality $d\to\infty$), 
while $\Sigma_{\bf p}(i\varepsilon)$ is some momentum dependent part.  
This last contribution can be due either to electron 
interactions with some ``additional'' collective modes or order parameter 
fluctuations, or may be induced by similar non--local contributions within 
the Hubbard model itself. No doublecounting problem arises in this approach,
as discussed in details in Ref. \cite{PRB05}. At the same time our procedure
as stressed in Refs. \cite{JTL05,PRB05,FNT06} does not represent any systematic
1/d expansion. Basic assumption here is the neglect of all 
interference processes of the local Hubbard interaction and non-local 
contributions owing to these additional scatterings (non-crossing approximation 
for appropriate diagrams) \cite{PRB05}, as illustrated by diagrams in
Fig. \ref{dDMFT_PG}.

The self--consistency equations of generalized DMFT+$\Sigma_{\bf p}$ 
approach are formulated as follows \cite{JTL05,PRB05}:
\begin{enumerate}
\item{Start with some initial guess of {\em local} self--energy
$\Sigma(i\varepsilon)$, e.g. $\Sigma(i\varepsilon)=0$}.  
\item{Construct $\Sigma_{\bf p}(i\varepsilon)$ within some (approximate) 
scheme, taking into account interactions with collective modes or order 
parameter fluctuations which in general can depend on $\Sigma(i\omega)$ and 
$\mu$.} 
\item{Calculate the local Green function 
\begin{equation} 
G_{ii}(i\varepsilon)=\frac{1}{N}\sum_{\bf p}\frac{1}{i\varepsilon+\mu
-\varepsilon({\bf p})-\Sigma(i\varepsilon)-\Sigma_{\bf p}(i\varepsilon)}.
\label{Gloc}
\end{equation}
}
\item{Define the ``Weiss field''
\begin{equation}
{\cal G}^{-1}_0(i\varepsilon)=\Sigma(i\varepsilon)+G^{-1}_{ii}(i\varepsilon).
\label{Wss}
\end{equation}
}
\item{Using some ``impurity solver'' calculate the single-particle Green 
function $G_d(i\varepsilon)$ for the effective Anderson impurity 
problem, placed at lattice site $i$, and defined by effective action
which is written, in obvious notations, as:
\begin{equation}
S_{\text{eff}}=-\int_{0}^{\beta}d\tau_1\int_{0}^{\beta}
d\tau_2c_{i\sigma}(\tau_1){\cal G}^{-1}_0(\tau_1-\tau_2)c^+_{i\sigma}(\tau_2)
+\int_{0}^{\beta}d\tau Un_{i\uparrow}(\tau)n_{i\downarrow}(\tau). 
\label{Seff}
\end{equation}
}
\item{Define a {\em new} local self--energy
\begin{equation}
\Sigma(i\omega)={\cal G}^{-1}_0(i\omega)-
G^{-1}_{d}(i\omega).
\label{StS}
\end{equation}
}
\item{Using this self--energy as 
``initial'' one in step 1, continue the procedure until (and if) convergency 
is reached to obtain 
\begin{equation} 
G_{ii}(i\varepsilon)=G_{d}(i\varepsilon).  
\label{G00}
\end{equation}
}
\end{enumerate}
Eventually, we get the desired Green function in the form of (\ref{Gk}),
where $\Sigma(i\varepsilon)$ and $\Sigma_{\bf p}(i\varepsilon)$ are those 
appearing at the end of our iteration procedure.

\section{Optical conductivity in DMFT+$\Sigma_{\bf p}$}
\label{opt_cond}

\subsection{Basic expressions for optical conductivity}

To calculate dynamic conductivity we use the general expression relating it
to retarded density--density correlation function $\chi^R(\omega,{\bf q})$
\cite{VW,Diagr}:
\begin{equation}
\sigma(\omega)=-\lim_{q\to 0}\frac{ie^2\omega}{q^2}\chi^R(\omega,{\bf q}),
\label{cond_op}
\end{equation}
where $e$ is electronic charge.

\begin{figure}
\includegraphics[clip=true,width=0.8\textwidth]{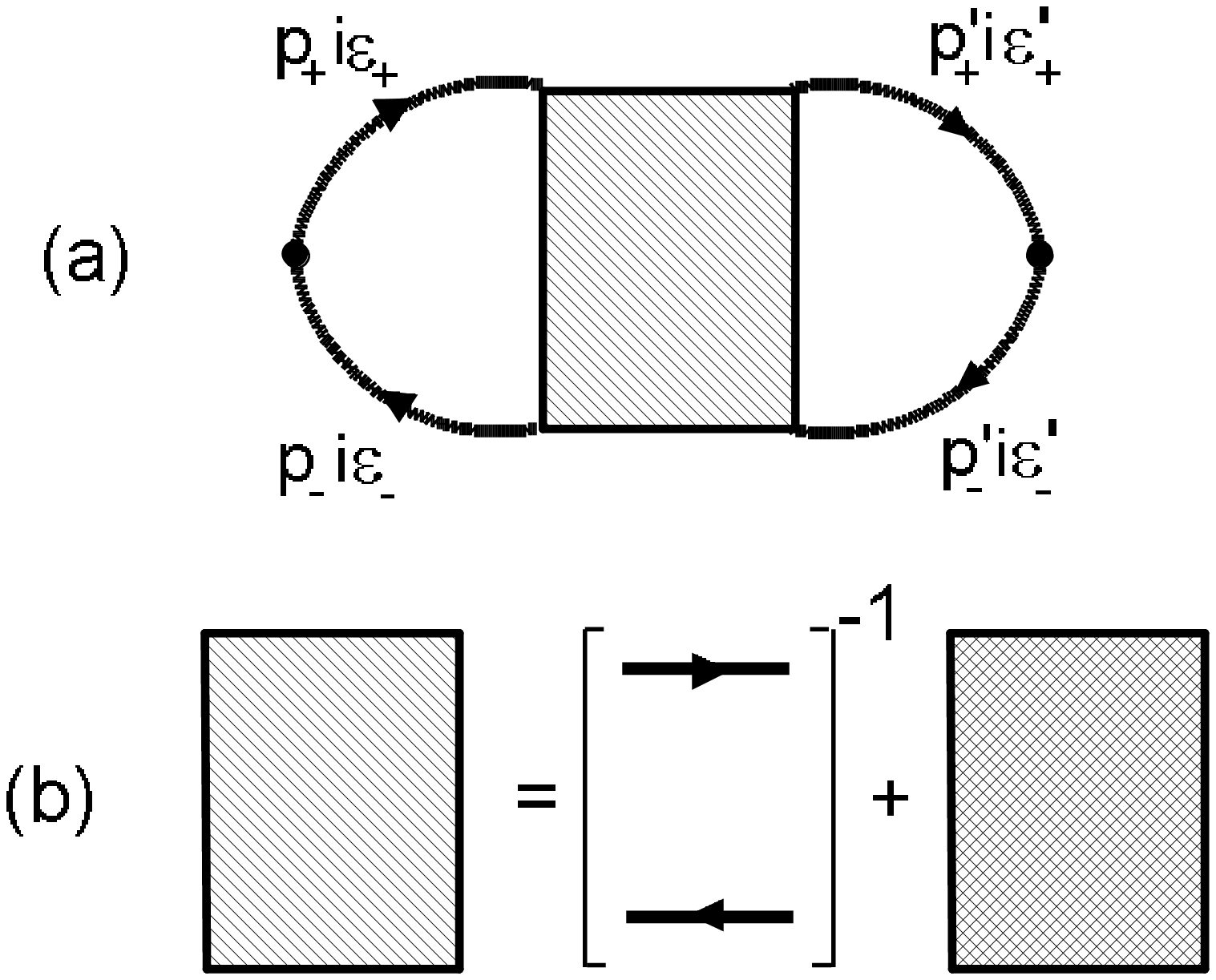}
\caption{Full polarization loop (a) with vertex part, which
includes free--electron contribution in addition to the standard vertex, 
containing all interactions (b). 
Here ${\bf p}_{\pm}={\bf p}\pm\frac{\bf q}{2}$, $\varepsilon_{\pm}=
\varepsilon\pm\frac{\omega}{2}$. }
\label{loop}
\end{figure}

Consider full polarization loop graph in Matsubara representation 
shown in Fig. \ref{loop}(a), which is conveniently  
(with explicit frequency summation) written as:
\begin{equation}
\Phi(i\omega,{\bf q})=\sum_{\varepsilon\varepsilon'}
\Phi_{i\varepsilon i\varepsilon'}(i\omega,{\bf q})\equiv\sum_{\varepsilon}
\Phi_{i\varepsilon}(i\omega,{\bf q})
\label{PhiM}
\end{equation}
and contains all possible interactions of our model, described by the full
vertex part of Fig. \ref{loop}(b). 
Note that, we use slightly unusual definition of
the vertex part to include the loop contribution without vertex corrections,
which shortens further diagrammatic expressions.
Retarded density--density correlation 
function is determined by appropriate analytic continuation of this loop and 
can be written as:
\begin{equation}
\chi^R(\omega,{\bf q})=
\int_{-\infty}^{\infty}\frac{d\varepsilon}{2\pi i}\left\{\left[f(\varepsilon_+) 
-f(\varepsilon_-)\right]\Phi_{\varepsilon}^{RA}({\bf q},\omega)
+f(\varepsilon_-)\Phi_{\varepsilon}^{RR}({\bf q},\omega)
-f(\varepsilon_+)\Phi_{\varepsilon}^{AA}({\bf q},\omega)\right\}
\label{cond_gen},
\end{equation}
where $f(\varepsilon)$ -- Fermi distribution,
$\varepsilon_{\pm}=\varepsilon\pm\frac{\omega}{2}$, while two--particle loops  
$\Phi_{\varepsilon}^{RA}({\bf q},\omega)$, 
$\Phi_{\varepsilon}^{RR}({\bf q},\omega)$, 
$\Phi_{\varepsilon}^{AA}({\bf q},\omega)$ are determined by appropriate
analytic continuations $(i\varepsilon+i\omega\to\varepsilon+\omega+i\delta, 
i\varepsilon\to\varepsilon\pm i\delta, \delta\to +0)$ in (\ref{PhiM}).
Then we can conveniently write dynamic conductivity as:
\begin{eqnarray}
\sigma(\omega)=\lim_{q\to 0}\left(-\frac{e^2\omega}{2\pi q^2}\right)
\int_{-\infty}^{\infty}d\varepsilon\left\{\left[f(\varepsilon_+)
-f(\varepsilon_-)\right]\left[\Phi_{\varepsilon}^{RA}({\bf q},\omega)
-\Phi_{\varepsilon}^{RA}(0,\omega)\right]+\right.\nonumber\\
\left.+f(\varepsilon_-)\left[\Phi_{\varepsilon}^{RR}({\bf q},\omega)
-\Phi_{\varepsilon}^{RR}(0,\omega)\right]-
f(\varepsilon_+)\left[\Phi_{\varepsilon}^{AA}({\bf q},\omega)-
\Phi_{\varepsilon}^{AA}(0,\omega)\right]\right\},
\label{cond_gener}
\end{eqnarray}
where the total contribution of additional terms with zero $q$ can be shown 
(with the use of general Ward identities \cite{Janis}) to be zero. 

\begin{figure}
\includegraphics[clip=true,width=0.8\textwidth]{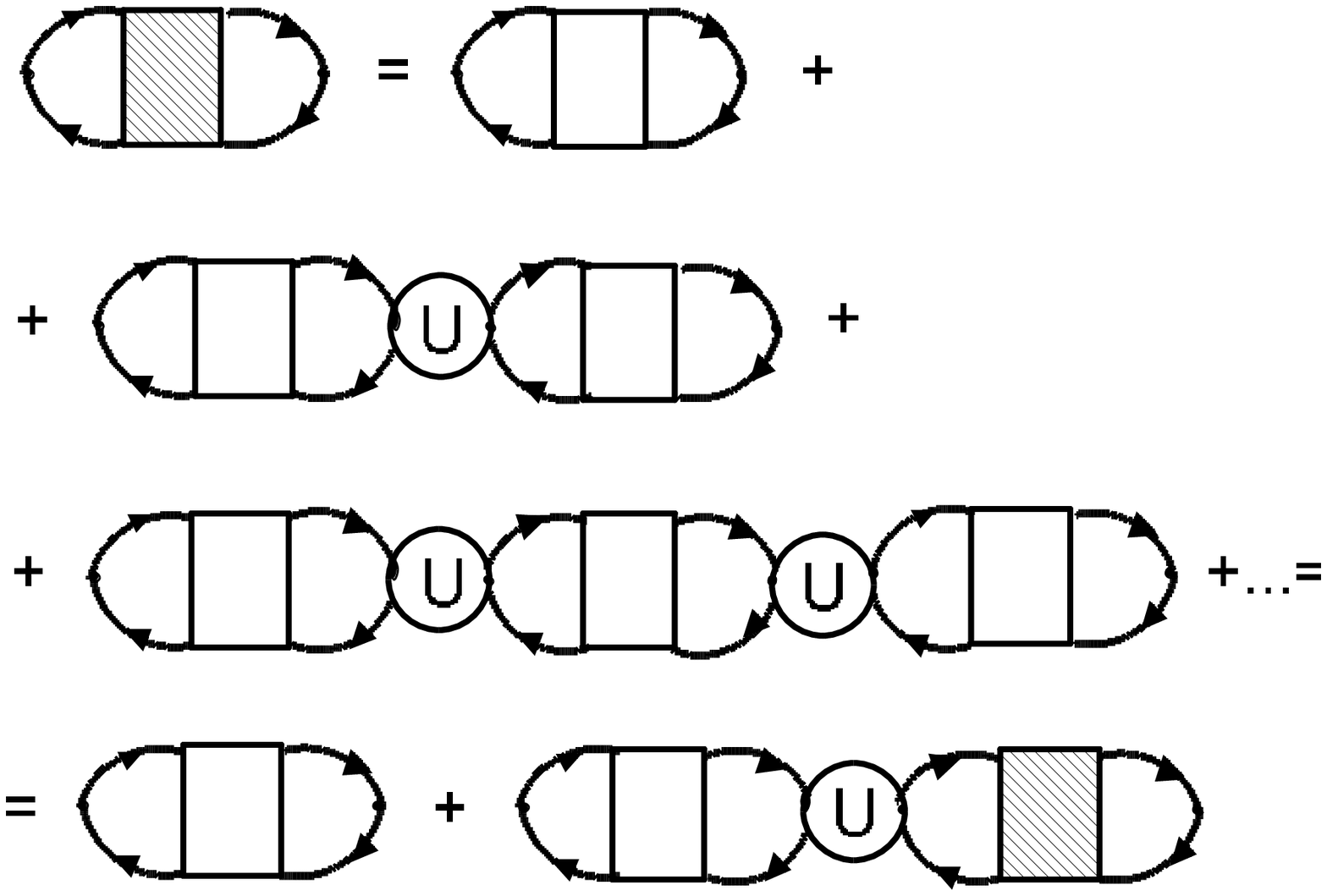}
\caption{Bethe--Salpeter equation for polarization loop in
DMFT+$\Sigma_{\bf p}$ approach. Circle represents irreducible vertex part
of DMFT, which contains only local interactions, surviving in the limit of
$d\to\infty$. Unshaded rectangular vertex represents non-local
interactions, e.g. with SDW (pseudogap) fluctuations, which is defined similarly
to Fig. \ref{loop}(b).}
\label{BS_loop}
\end{figure}

To calculate $\Phi_{i\varepsilon i\varepsilon'}(i\omega,{\bf q})$,
entering the sum over Matsubara frequencies in (\ref{PhiM}), 
in DMFT+$\Sigma_{\bf p}$ approximation, which
neglects interference between local Hubbard interaction and non-local 
contributions due to additional scatterings, e.g. by SDW pseudogap 
fluctuations \cite{PRB05}, we can write down Bethe--Salpeter equation,
shown diagrammatically in Fig. \ref{BS_loop}, where we have introduced
irreducible (local) vertex $U_{i\varepsilon i\varepsilon'}(i\omega)$ 
of DMFT and ``rectangular'' vertex, defined
as in Fig. \ref{loop}(b) and containing all interactions with fluctuations.
Analytically this equation can be written as:
\begin{equation}
\Phi_{i\varepsilon i\varepsilon'}(i\omega,{\bf q})=
\Phi^0_{i\varepsilon}(i\omega,{\bf q})
\delta_{\varepsilon\varepsilon'}+
\Phi^0_{i\varepsilon}(i\omega,{\bf q})\sum_{\varepsilon''}
U_{i\varepsilon i\varepsilon''}(i\omega)
\Phi_{i\varepsilon'' i\varepsilon'}(i\omega,{\bf q}),
\label{BS_equ}
\end{equation}
where $\Phi^0_{i\varepsilon}(i\omega,{\bf q})$ is the desired
function calculated neglecting vertex corrections due to Hubbard interaction 
(but taking into account all non-local interactions with fluctuations, 
considered here to be static).
Note that all $q$-dependence here is determined by 
$\Phi^0_{i\varepsilon}(i\omega,{\bf q})$ as the vertex 
$U_{i\varepsilon i\varepsilon'}(i\omega)$ is local and $q$-independent. 

\begin{figure}
\includegraphics[clip=true,width=0.8\textwidth]{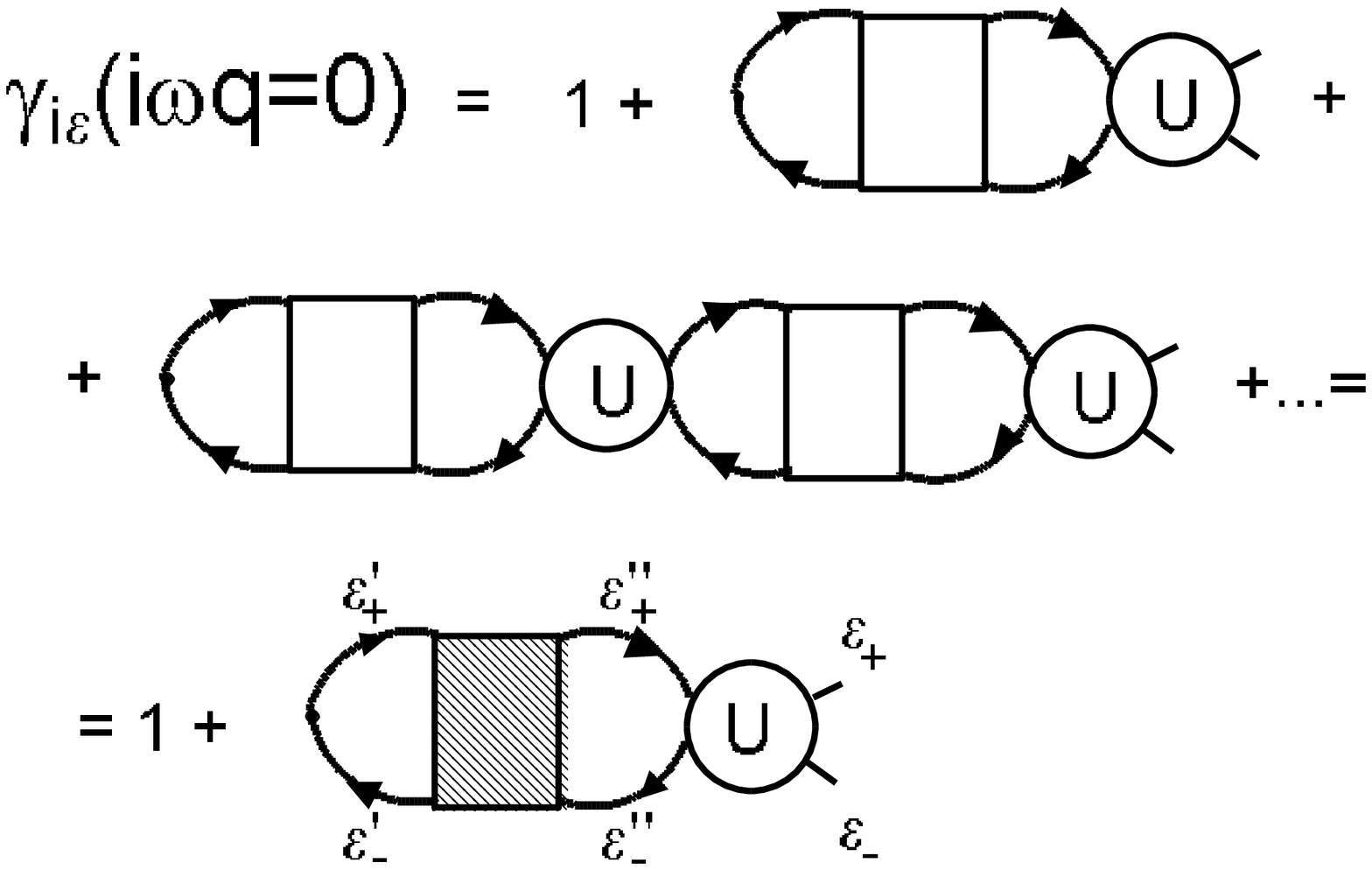}
\caption{Effective vertex $\gamma_{i\varepsilon}(i\omega,{\bf q}=0)$ used in calculations of
conductivity.}
\label{gamma}
\end{figure}

As is clear from (\ref{cond_gener}), to calculate conductivity, we need only
to find $q^2$-contribution to 
$\Phi(i\omega,{\bf q})$ defined in (\ref{PhiM}). This can be done in the
following way. First of all, note that all the loops in (\ref{BS_equ}) contain
$q$-dependence starting from terms of the order of $q^2$. Then we can take
an arbitrary loop (crossection) in the expansion of (\ref{BS_equ}) (see
Fig. \ref{BS_loop}), calculating it up to terms of the order of $q^2$, and 
make resummation of all contributions to the right and to the left from 
this crossection (using the obvious left -- right symmetry of diagram
summation in Bethe -- Salpeter equation), putting ${\bf q}=0$ in all these 
graphs. This is equivalent to simple $q^2$-differentiation of the expanded 
version of Eq. (\ref{BS_equ}).
This procedure immediately leads to the following relation for 
$q^2$-contribution to (\ref{PhiM}):
\begin{equation}
\phi(i\omega)\equiv\lim_{q\to 0}\frac{\Phi(i\omega,{\bf q})-
\Phi(i\omega,0)}{q^2}=\sum_{\varepsilon}\gamma^2_{i\varepsilon}(i\omega,{\bf q}=0)
\phi^0_{i\varepsilon}(i\omega)
\label{fi_func}
\end{equation}
where
\begin{equation}
\phi^0_{i\varepsilon}(i\omega)\equiv\lim_{q\to 0}
\frac{\Phi^0_{i\varepsilon}(i\omega,{\bf q})-
\Phi^0_{i\varepsilon}(i\omega,0)}{q^2}
\label{fi0_func}
\end{equation}
with $\Phi^0_{i\varepsilon}(i\omega,{\bf q})$
containing vertex corrections only due to non-local
(pseudogap) fluctuations, while one-particle Green's functions in it are taken
with self-energies due to both these fluctuations and local DMFT-like 
interaction, like in Eq. (\ref{Gk}). The vertex 
$\gamma_{i\varepsilon}(i\omega,{\bf q}=0)$ 
is determined diagrammatically as shown in Fig. \ref{gamma}, or analytically:
\begin{equation}
\gamma_{i\varepsilon}(i\omega,{\bf q}=0)=
1+\sum_{\varepsilon'\varepsilon''}
U_{i\varepsilon i\varepsilon''}(i\omega)
\Phi_{i\varepsilon''i\varepsilon'}(i\omega,{\bf q}=0).
\label{gamma_equ}
\end{equation}
Now using Bethe--Salpeter equation (\ref{BS_equ}) we can write explicitly:
\begin{eqnarray}
\gamma_{i\varepsilon}(i\omega,{\bf q}=0)=1+\sum_{\varepsilon'}
\frac{\Phi_{i\varepsilon i\varepsilon'}(i\omega,{\bf q}=0)-
\Phi^0_{i\varepsilon}(i\omega,{\bf q}=0)}{\Phi^0_{i\varepsilon}(i\omega,{\bf q}=0)}=
\nonumber\\
=\frac{\sum_{\varepsilon'}\Phi_{i\varepsilon i\varepsilon'}(i\omega,{\bf q}=0)}
{\Phi^0_{i\varepsilon}(i\omega,{\bf q}=0)}.
\label{gamm_expl} 
\end{eqnarray}
For ${\bf q}=0$ we have the following Ward identity, which can be obtained by direct
generalization of the proof given in Refs. \cite{VW,Janis} (see Appendix \ref{A}):
\begin{equation}
(-i\omega)\Phi_{i\varepsilon}(i\omega,{\bf q}=0)=
(-i\omega)\sum_{\varepsilon'}\Phi_{i\varepsilon i\varepsilon'}(i\omega,{\bf q}=0)=
\sum_{\bf p}G(i\varepsilon+i\omega,{\bf p})-
\sum_{\bf p}G(i\varepsilon,{\bf p}).
\label{Ward1}
\end{equation}
Denominator of (\ref{gamm_expl}) contains vertex corrections only from
non-local correlations (e.g. pseudogap fluctuations), while Green's functions
here are ``dressed'' {\em both} by these correlations and local
(DMFT) Hubbard interaction. Thus we may consider the loop entering the
denominator as dressed by (pseudogap) fluctuations only, but with ``bare''
Green's functions:
\begin{equation}
\tilde G_{0}(i\varepsilon,{\bf p})=\frac{1}{i\varepsilon+\mu
-\varepsilon({\bf p})- \Sigma(i\varepsilon)},
\label{G_DMFT} 
\end{equation}
where $\Sigma(i\varepsilon)$ is local contribution to self-energy from DMFT.
For this problem we have the following Ward identity, similar to (\ref{Ward1}) 
(see Appendix \ref{A}):
\begin{eqnarray}
\sum_{\bf p}G(i\varepsilon+i\omega,{\bf p})-\sum_{\bf p}G(i\varepsilon,{\bf p})
=\Phi_{i\varepsilon}^0(i\omega,{\bf q}=0)\left[\Sigma(i\varepsilon+i\omega)-
\Sigma(i\varepsilon)-i\omega\right]\equiv\nonumber\\
\equiv\Phi_{i\varepsilon}^0(i\omega,{\bf q}=0)\left[\Delta\Sigma(i\omega)-i\omega\right],
\label{Ward2}
\end{eqnarray}
where we have introduced 
\begin{equation}
\Delta\Sigma(i\omega)=\Sigma(i\varepsilon+i\omega)-
\Sigma(i\varepsilon).
\label{D_Sigma}
\end{equation}
Thus, using (\ref{Ward1}), (\ref{Ward2}) in (\ref{gamm_expl}) we get 
the final expression for $\gamma_{i\varepsilon}(i\omega,{\bf q}=0)$:  
\begin{equation} 
\gamma_{i\varepsilon}(i\omega,{\bf q}=0)=1-\frac{\Delta\Sigma(i\omega)}{i\omega}.
\label{gamm_fin}
\end{equation}
Then (\ref{fi_func}) reduces to:
\begin{equation}
\phi(i\omega)=\sum_{\varepsilon}\phi^0_{i\varepsilon}(i\omega)
\left[1-\frac{\Delta\Sigma(i\omega)}
{i\omega}\right]^2.
\label{fi_funct}
\end{equation}
Analytic continuation to real frequencies is obvious and using (\ref{fi_func}),
(\ref{fi_funct}) in (\ref{cond_gener}) we can write the final expression for 
the real part of dynamic conductivity as:
\begin{eqnarray}
{\rm{Re}}\sigma(\omega)=\frac{e^2\omega}{2\pi}
\int_{-\infty}^{\infty}d\varepsilon\left[f(\varepsilon_-)
-f(\varepsilon_+)\right]{\rm{Re}}\left\{\phi^{0RA}_{\varepsilon}(\omega)\left[1-
\frac{\Sigma^R(\varepsilon_+)-\Sigma^A(\varepsilon_-)}{\omega}\right]^2-
\right.\nonumber\\
\left.-\phi^{0RR}_{\varepsilon}(\omega)\left[1-
\frac{\Sigma^R(\varepsilon_+)-\Sigma^R(\varepsilon_-)}{\omega}\right]^2
\right\}.
\label{cond_final}
\end{eqnarray}
Thus we have achieved a great simplification of our problem. To calculate
optical conductivity in DMFT+$\Sigma_{\bf p}$ we only have to solve 
single--particle problem as described by DMFT+$\Sigma_{\bf p}$ procedure
above to determine self--consistent values of local self--energies 
$\Sigma(\varepsilon_{\pm})$, while non-trivial contribution of non-local
correlations are to be included via (\ref{fi0_func}), which is to be 
calculated in some approximation, taking into account only interaction with
non-local (e.g. pseudogap) fluctuations, but using the ``bare'' Green's
functions of the form (\ref{G_DMFT}), which include local self--energies
already determined in the general DMFT+$\Sigma_{\bf p}$ procedure.
Actually (\ref{cond_final}) provides also an effective algorithm to
calculate dynamic conductivity in standard DMFT (neglecting any non-local
correlations), as (\ref{fi0_func}) is then easily calculated from a simple
loop diagram, determined by two Green's functions and free {\em scalar} 
vertices.  As usual, there is no need to calculate vertex corrections within 
DMFT itself, as was proven first considering the loop with {\em vector} 
vertices \cite{pruschke,georges96}.

\subsection{Recurrence relations for self--energy and vertex--parts.}
\label{kselfvertex}

As we are mainly interested in the pseudogap state of copper oxides,
we shall further concentrate on the effects of scattering of electrons from 
collective short-range SDW--like antiferromagnetic spin  
fluctuations. In a kind of a simplified approach, valid only for high enough
temperatures \cite{Sch,KS} we shall calculate $\Sigma_{\bf p}(i\omega)$ for an 
electron moving in the quenched random field of (static) Gaussian spin  
fluctuations with dominant scattering momentum transfers from the 
vicinity of some characteristic vector ${\bf Q}$ (``hot-spots'' model 
\cite{MS}), using (as we have done in Refs. \cite{JTL05,PRB05,FNT06})
slightly generalized version of the recurrence procedure 
proposed in \cite{Sch,KS,MS79} (see also Ref. \cite{Diagr}), which takes into 
account {\em all} Feynman diagrams describing the scattering of electrons by 
this random field.
In general, neglect of fluctuation dynamics, overestimates pseudogap effects.
Referring the reader to earlier papers for details \cite{Sch,KS,JTL05,PRB05,FNT06}, here we just
start with the main recurrence relation, determining the self--energy:  
\begin{equation} 
\Sigma_{k}(i\varepsilon,{\bf p})=\Delta^2\frac{s(k)} 
{i\varepsilon+\mu-\Sigma(i\varepsilon) -\varepsilon_k({\bf p})
+inv_k\kappa-\Sigma_{k+1}(i\varepsilon,{\bf p})}\;\;.  
\label{rec} 
\end{equation}
Usually one takes the value of $\Sigma_{k+1}$ for large enough $k$ equal to zero
and doing recurrence backwards to $k=1$ gets the desired physical self--energy 
$\Sigma(i\varepsilon,{\bf p})=\Sigma_1(i\varepsilon,{\bf p})$ 
\cite{KS,MS79,Diagr}.
 
In Eq. (\ref{rec}) $\Delta$ characterizes the energy scale and
$\kappa=\xi^{-1}$ is the inverse correlation length of short--range
SDW fluctuations, $\varepsilon_k({\bf p})=\varepsilon({\bf p+Q})$ and 
$v_k=|v_{\bf p+Q}^{x}|+|v_{\bf p+Q}^{y}|$ 
for odd $k$ while $\varepsilon_k({\bf p})=\varepsilon({\bf p})$ and $v_{k}=
|v_{\bf p}^x|+|v_{\bf p}^{y}|$ for even $k$. The velocity projections
$v_{\bf p}^{x}$ and $v_{\bf p}^{y}$ are determined by usual momentum derivatives
of the ``bare'' electronic energy dispersion $\varepsilon({\bf p})$. Finally,
$s(k)$ represents a combinatorial factor, which here is always assumed to be
that corresponding to the case of Heisenberg spin fluctuations in  
``nearly antiferromagnetic Fermi--liquid'' 
(spin--fermion (SF) model of Ref.~\cite{Sch}, SDW-type fluctuations):
\begin{equation} 
s(k)=\left\{\begin{array}{cc}
\frac{k+2}{3} & \mbox{for odd $k$} \\
\frac{k}{3} & \mbox{for even $k$}.
\end{array} \right.
\label{vspin}
\end{equation}
As was stressed in Refs. \cite{PRB05,FNT06} this procedure introduces an 
important length scale $\xi$ not present in standard DMFT, which
mimics the effect of short--range (SDW) correlations within 
fermionic ``bath'' surrounding the DMFT effective single Anderson impurity.  

An important aspect of the theory is that both parameters $\Delta$ and 
$\xi$ can in principle be calculated from the microscopic model at hand
\cite{PRB05}, but here we consider these as phenomenological parameters of
the theory (e.g. to be determined from experiments). 

\begin{figure}
\includegraphics[clip=true,width=0.8\textwidth]{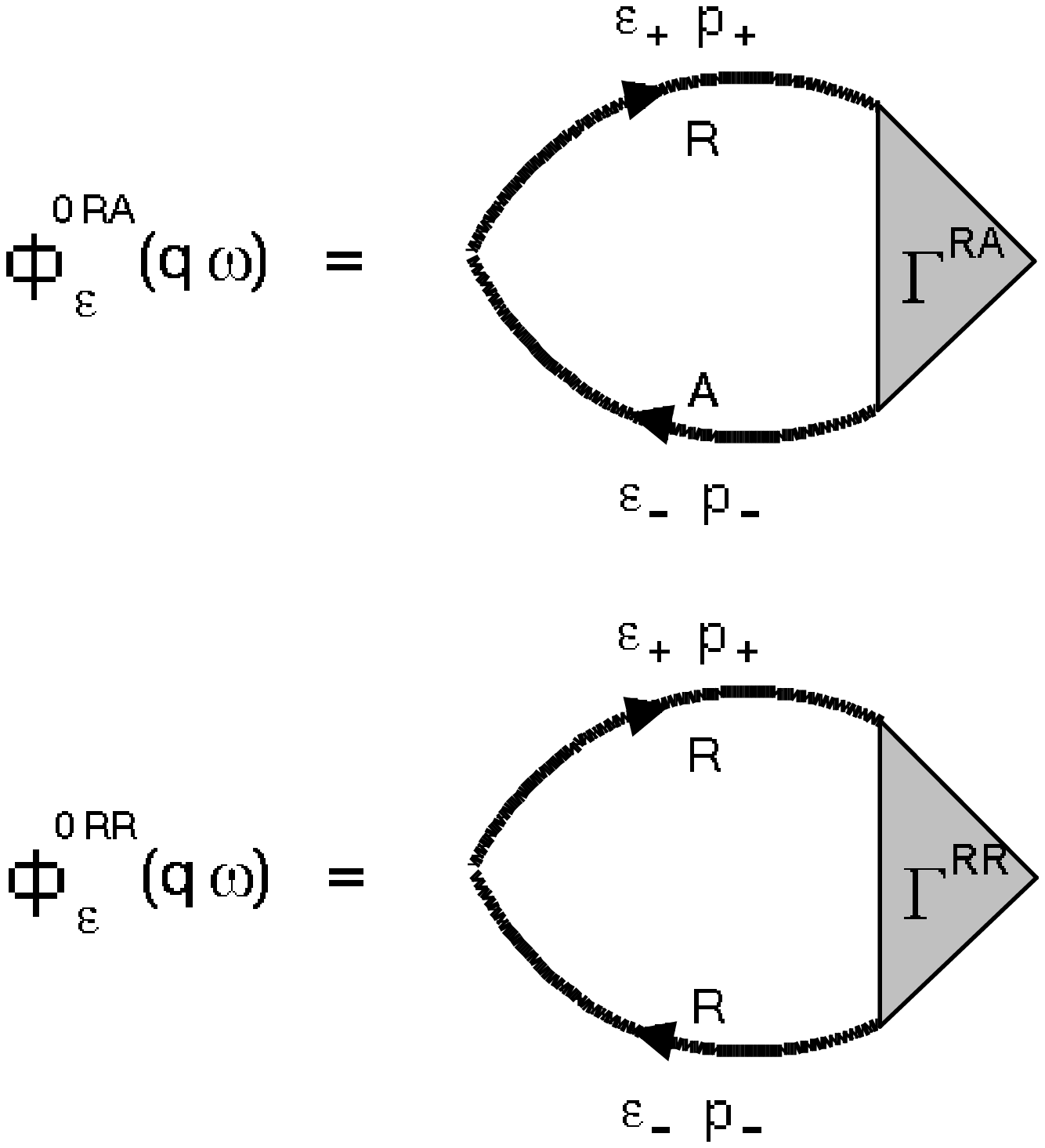}
\caption{Diagrammatic representation of $\Phi^{0RA}_{\varepsilon}(\omega,{\bf q})$.} 
\label{loop_fi} 
\end{figure}

\begin{figure}
\includegraphics[clip=true,width=0.8\textwidth]{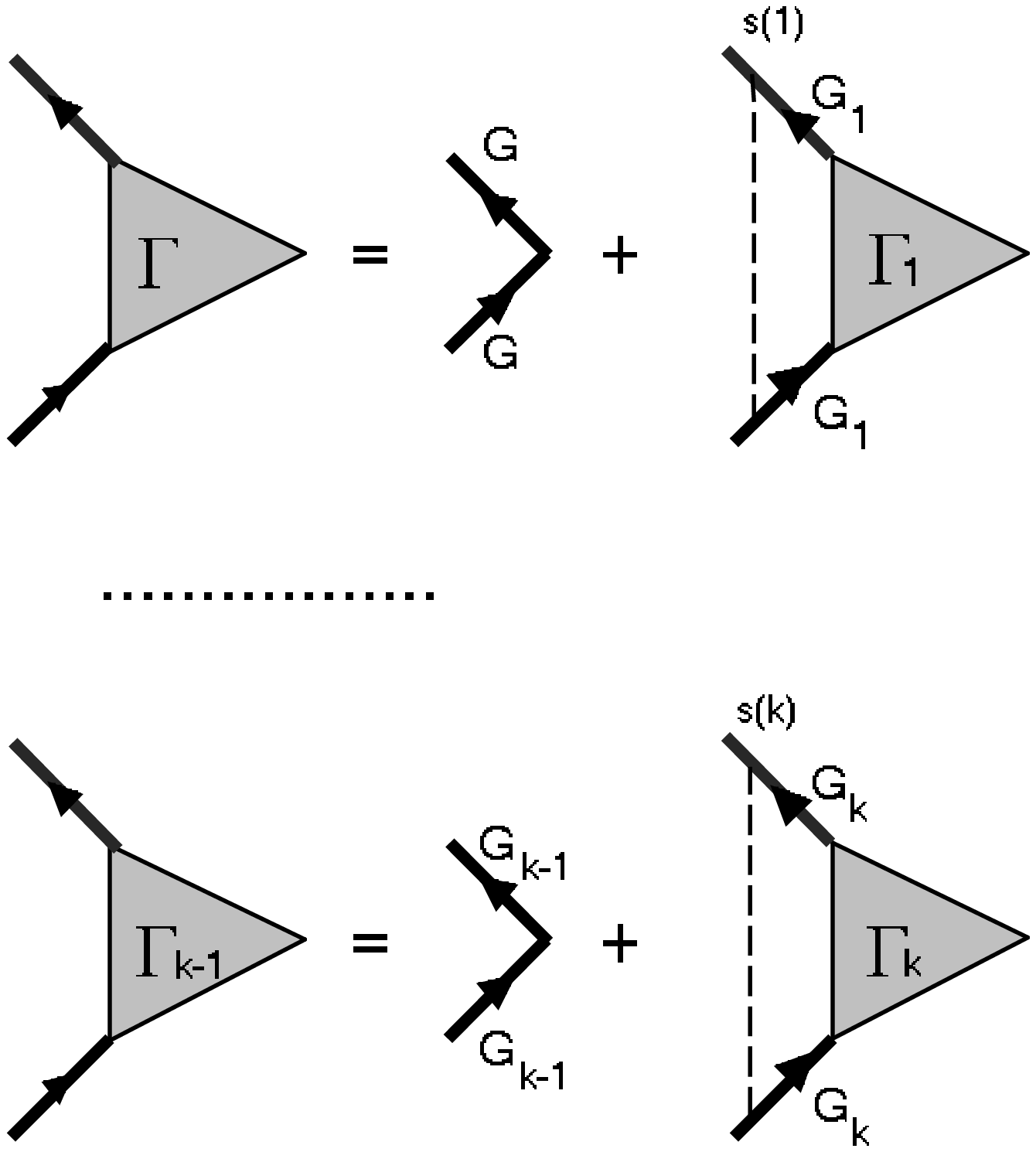}
\caption{Recurrence relations for the vertex part. Dashed lines denote 
$\Delta^2$.}
\label{recvertx}
\end{figure}

Now to calculate optical conductivity we need the knowledge of the 
basic block $\Phi^0_{i\varepsilon}(i\omega,{\bf q})$, entering 
(\ref{fi0_func}), or, more precisely, appropriate functions analytically 
continued to real frequencies: $\Phi^{0RA}_{\varepsilon}(\omega,{\bf q})$ and
$\Phi^{0RR}_{\varepsilon}(\omega,{\bf q})$, which in turn define
$\phi^{0RA}_{\varepsilon}(\omega)$ and $\phi^{0RR}_{\varepsilon}(\omega)$
entering (\ref{cond_final}), and defined by obvious relations similar to
(\ref{fi0_func}): 
\begin{equation}
\phi^{0RA}_{\varepsilon}(\omega)=\lim_{q\to 0}
\frac{\Phi^{0RA}_{\varepsilon}(\omega,{\bf q})-
\Phi^{0RA}_{\varepsilon}(\omega,0)}{q^2},
\label{fi0RA_func}
\end{equation}
\begin{equation}
\phi^{0RR}_{\varepsilon}(\omega)=\lim_{q\to 0}
\frac{\Phi^{0RR}_{\varepsilon}(\omega,{\bf q})-
\Phi^{0RR}_{\varepsilon}(\omega,0)}{q^2}
\label{fi0RR_func}.
\end{equation}
By definition we have:
\begin{equation}
\Phi^{0RA}_{\varepsilon}(\omega,{\bf q})=\sum_{\bf p}
G^R(\varepsilon_+,{\bf p_+})G^A(\varepsilon_-,{\bf p_-})
\Gamma^{RA}(\varepsilon_-,{\bf p}_-;\varepsilon_+,{\bf p}_+)
\label{PhiRA}
\end{equation}
\begin{equation}
\Phi^{0RR}_{\varepsilon}(\omega,{\bf q})=\sum_{\bf p}
G^R(\varepsilon_+,{\bf p_+})G^R(\varepsilon_-,{\bf p_-})
\Gamma^{RR}(\varepsilon_-,{\bf p}_-;\varepsilon_+,{\bf p}_+)\\ \nonumber,
\label{PhiRR}
\end{equation}
which are shown diagrammatically in Fig. \ref{loop_fi}. 
Here Green's functions $G^R(\varepsilon_+,{\bf p_+})$ and
$G^A(\varepsilon_-,{\bf p_-})$ are defined by analytic continuation 
$(i\varepsilon\to\varepsilon\pm i\delta)$ of Matsubara Green's functions 
(\ref{Gk}) determined by recurrence procedure (\ref{rec}), while vertices
$\Gamma^{RA}(\varepsilon_-,{\bf p}_-;\varepsilon_+,{\bf p}_+)$ and
$\Gamma^{RR}(\varepsilon_-,{\bf p}_-;\varepsilon_+,{\bf p}_+)$, containing
all vertex corrections due to pseudogap fluctuations are given by the
recurrence procedure, derived first (for one-dimensional case) in Ref.
\cite{ST91} (see also Ref. \cite{Diagr}) and generalized for two--dimensional
problem in Ref. \cite{SS02} (see also Ref. \cite{Sch}). The basic idea used here
is that an arbitrary diagram for the vertex part can be obtained by an insertion 
of an ``external field'' line into the appropriate diagram for the self--energy 
\cite{S74,ST91,SS02}. In our model we can limit ourselves only to diagrams with 
non--intersecting interaction lines with additional combinatorial factors 
$s(k)$ in ``initial'' interaction vertices \cite{MS79,Sch,KS}. 
Thus, all diagrams for the vertex
part are, in fact, generated by simple ladder diagrams with additional 
$s(k)$--factors, associated with interaction lines \cite{ST91,SS02} 
(see also \cite{Diagr}).
Then we obtain the system of recurrence relations for the vertex part
$\Gamma^{RA}(\varepsilon_-,{\bf p}_-;\varepsilon_+,{\bf p}_+)$
shown by diagrams  of Fig. \ref{recvertx}. Analytically it  
has the following form \cite{SS02}, where we now also included contributions due
to local (DMFT) self-energies, originating from DMFT+$\Sigma_p$ loop:  
\begin{eqnarray}
&& {\Gamma}_{k-1}^{RA}(\varepsilon_-,{\bf p}_-;\varepsilon_+,{\bf p}_+)=
1+\Delta^2s(k)G_{k}^A(\varepsilon_-,{\bf p_-})
G_{k}^R(\varepsilon_+,{\bf p_+})\times\nonumber\\
&& \times\left\{1+ 
\frac{2iv_k\kappa k}{\omega-\varepsilon_k({\bf p}_+)
+\varepsilon_k({\bf p}_-)-\Sigma^R(\varepsilon_+)+\Sigma^A(\varepsilon_-)
-\Sigma_{k+1}^R(\varepsilon_+,{\bf p_+})
+\Sigma^A_{k+1}(\varepsilon_-,{\bf p_-})}\right\}\times\nonumber\\
&& \times{\Gamma}_k^{RA}(\varepsilon_-,{\bf p}_-;\varepsilon_+,{\bf p}_+),
\label{JrecRA}
\end{eqnarray}
and
\begin{equation}
G^{R,A}_{k}(\varepsilon_{\pm},{\bf p}_{\pm})=\frac{1}{\varepsilon_{\pm}-
\varepsilon_k({\bf p}_{\pm}) \pm ikv_k\kappa-\Sigma^{R,A}(\varepsilon_{\pm})
-\Sigma^{R,A}_{k+1}(\varepsilon_{\pm},{\bf p}_{\pm})}.
\label{G}
\end{equation}
The ``physical'' vertex 
$\Gamma^{RA}(\varepsilon_-,{\bf p}_-;\varepsilon_+,{\bf p}_+)$
is determined as $\Gamma^{RA}_{k=0}(\varepsilon_-,{\bf p}_-;\varepsilon_+,
{\bf p}_+)$. Recurrence procedure (\ref{JrecRA}) takes into account
{\em all} perturbation theory diagrams for the vertex part.
For $\kappa\to 0\quad (\xi\to\infty)$ (\ref{JrecRA}) reduces to the series
studied in Ref. \cite{S74} (cf. also Ref. \cite{Sch}), which can be summed
exactly in analytic form. Standard ``ladder'' approximation corresponds in
our scheme to the case of combinatorial factors $s(k)$ in (\ref{JrecRA}) being
equal to 1 \cite{ST91}.

Recurrence procedure for 
$\Gamma^{RR}(\varepsilon_-,{\bf p}_+;\varepsilon_+,{\bf p}_+)$ differs from
(\ref{JrecRA}) only by obvious replacements $A\to R$ and the whole expression
in figure brackets in the r.h.s. of (\ref{JrecRA}) just replaced by 1:
\begin{eqnarray}
{\Gamma}_{k-1}^{RR}(\varepsilon_-,{\bf p}_-;\varepsilon_+,{\bf p}_+)=
1+\Delta^2s(k)G_{k}^R(\varepsilon_-,{\bf p_-})
G_{k}^R(\varepsilon_+,{\bf p_+})
{\Gamma}_k^{RR}(\varepsilon_-,{\bf p};\varepsilon_+,{\bf p}_+).
\label{JrecRR}
\nonumber\\
\end{eqnarray}
Note that DMFT (Hubbard) interaction enters these equations only via 
local self-energies $\Sigma^{R,A}(\varepsilon_{\pm})$ calculated 
self-consistently according to our DMFT+$\Sigma_p$ procedure.

Equations (\ref{Gk}), (\ref{rec}), (\ref{JrecRA}), (\ref{JrecRR}) together 
with (\ref{fi0RA_func}), (\ref{fi0RR_func}) and (\ref{cond_final}) provide us 
with the complete self-consistent procedure to calculate optical conductivity 
of our model in DMFT+$\Sigma_{\bf p}$ approach.

\section{Results and discussion}
\label{results}

\subsection{Generalities}

In the following, we shall discuss our results for a standard one-band
Hubbard model on a square lattice. The ``bare'' electronic dispersion in 
tight-binding approximation, with the account of nearest ($t$) and
next nearest ($t'$) neighbour hoppings, is given by:
\begin{equation}
\varepsilon({\bf p})=-2t(\cos p_xa+\cos p_ya)-4t'\cos p_xa\cos p_ya\;\;,
\label{spectr}
\end{equation}
where $a$ is the lattice constant. To be concrete, below we present results for
$t=0.25$eV (more or less typical for cuprates) 
and $t'/t$=-0.4 (which gives Fermi surface similar to those observed 
in many cuprates). 

For the square lattice the bare bandwidth is $W=8t$.
To study strongly correlated metallic state obtained as doped Mott insulator
we have used the value for the Hubbard interaction $U=40t$ and 
filling factors $n=1.0$ (half-filling) and $n=0.8$ (hole doping). 
For correlated metal with $W\gtrsim U$ we have taken typical
values like $U=4t$, $U=6t$ and $U=10t$ for $U\gtrsim W$. 
Calculations were performed for different 
fillings: half-filling ($n=1.0$) and for $n=0.8,0.9$ (hole doping). 
As typical values for $\Delta$ we have chosen 
$\Delta=t$ and  $\Delta=2t$ and for correlation length $\xi=2a$ and 
$\xi=10a$ (motivated mainly by experimental data for cuprates~\cite{MS,Sch}).

To solve an effective Anderson impurity problem of DMFT we applied
a reliable numerically exact method of numerical renormalization group (NRG) 
\cite{NRG,BPH}, which, actually, allowed us to work with real frequencies from
the very beginning, overcoming possible difficulties of performing
analytical continuation numerically. Calculations were performed for 
two different temperatures: $T=0.088t$ and $T=0.356t$.

All necessary integrations were done directly, e.g. over the whole
Brillouin zone (with the account of obvious symmetries), or wide enough 
frequency range. Integration momenta are made dimensionless in a natural way 
with the help of the lattice constant $a$. Conductivity is measured in 
units of the universal conductivity in two--dimensions: 
$\sigma_0=\frac{e^2}{\hbar}=2.5\ 10^{-4}$ Ohm$^{-1}$.

\subsection{Optical conductivity in standard DMFT}

Optical conductivity was calculated for different combinations of parameters 
of the model. Below we present only a fraction of our results, which are,
probably, most relevant for copper oxides.
We shall start with presenting some typical results, obtained within our
formalism in conventional DMFT approximation, neglecting pseudogap fluctuations,
just to introduce the basic physical picture and demonstrate the effectiveness
of our approach.

\begin{figure}
\includegraphics[clip=true,width=1.0\textwidth]{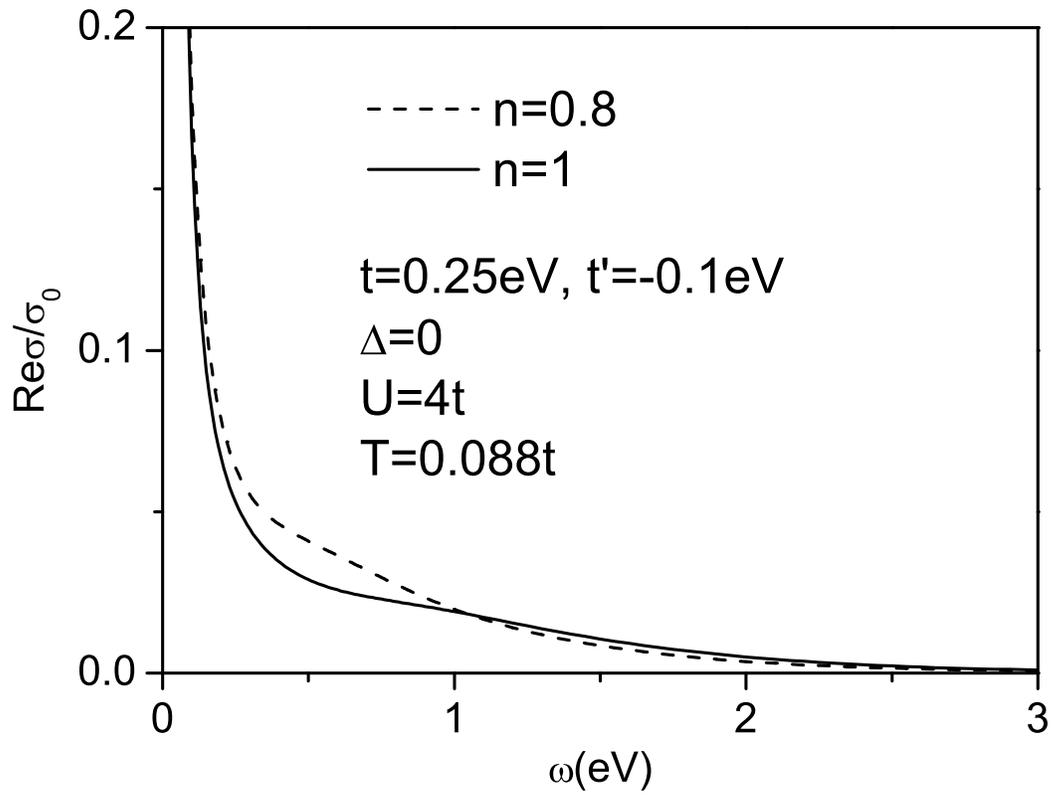}
\caption{Real part of optical conductivity for correlated metal ($U=4t$,
$t'=-0.4t$, $t=0.25$ eV) in 
DMFT approximation for two values of filling factor: $n=1$ and $n=0.8$.
Temperature $T=0.088t$.} 
\label{DMFT} 
\end{figure}

Characteristic feature of the strongly correlated metallic state
is the coexistence of lower and upper Hubbard bands splitted by the value of 
$\sim U$ with a quasiparticle peak at the Fermi level 
\cite{pruschke,georges96}.
For the case of strongly correlated metal with $W\gtrsim U$ we observe
almost no contribution from excitations to upper Hubbard model in optical
conductivity, as can be seen in Fig. \ref{DMFT} (where we show the real
part of conductivity Re$\sigma(\omega)$). This contribution is almost completely
masked by a typical Drude--like frequency behavior, with only slightly
non--monotonous behavior for $\omega\sim U$, which completely
disappears as we rise the temperature.

\begin{figure}
\includegraphics[clip=true,width=1.0\textwidth]{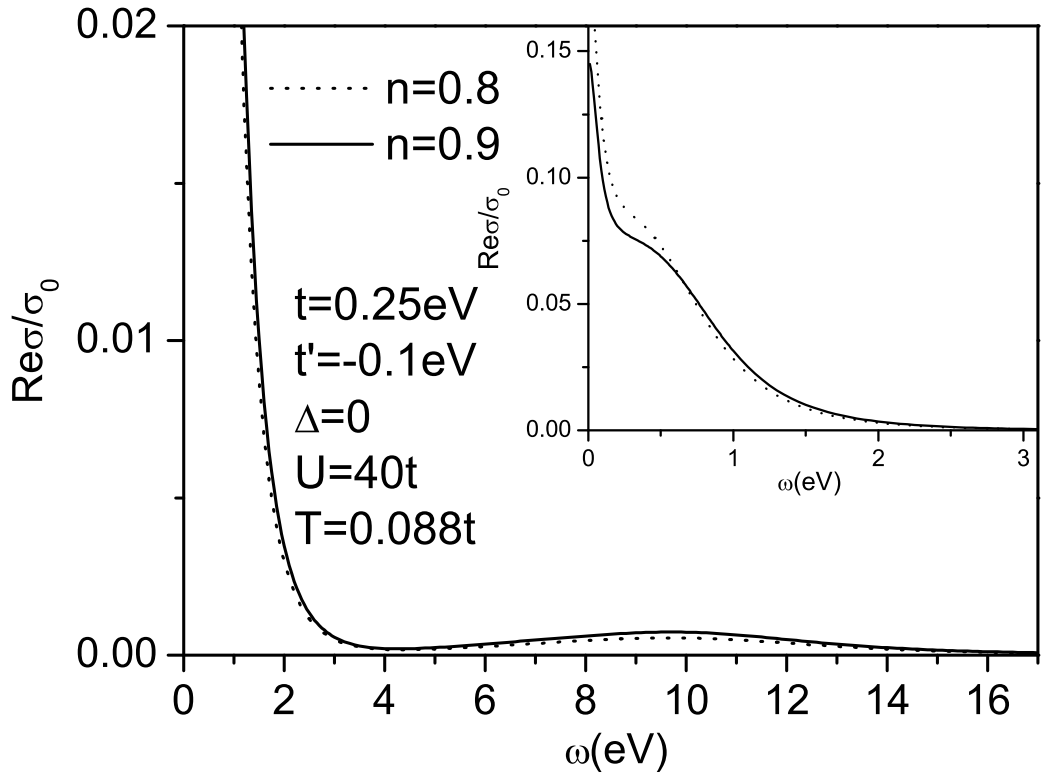}
\caption{Real part of optical conductivity for doped Mott insulator ($U=40t$,
$t'=-0.4t$, $t=0.25$ eV) 
in DMFT approximation. Filling factors are: $n=0.8$ and $n=0.9$, temperature
$T=0.088t$. Small frequency behavior is shown in more details at
the insert.} 
\label{DMFT_i} 
\end{figure}

Situation is different in doped Mott insulator with $U\gg W$. 
In Fig. \ref{DMFT_i} we clearly observe an additional maximum of optical
absorption for $\omega\sim U$, however, at smaller frequencies we again
observe typical Drude--like behavior, slightly non--monotonous for small
frequencies due to quasiparticle band formation 
(see insert in Fig. \ref{DMFT_i}).

These and similar results are more or less well known from the previous 
studies \cite{pruschke,georges96}, and are quoted here only to demonstrate
the consistency of our formalism and to prepare the reader for new results,
showing pseudogap behavior.

\subsection{Optical conductivity in DMFT+$\Sigma_{\bf p}$}

\subsubsection{Correlated metal}

Let us start the discussion of results obtained 
within our generalized DMFT+$\Sigma_{\bf p}$ approach
for the case of $W\gtrsim U$.

\begin{figure}
\includegraphics[clip=true,width=1.0\textwidth]{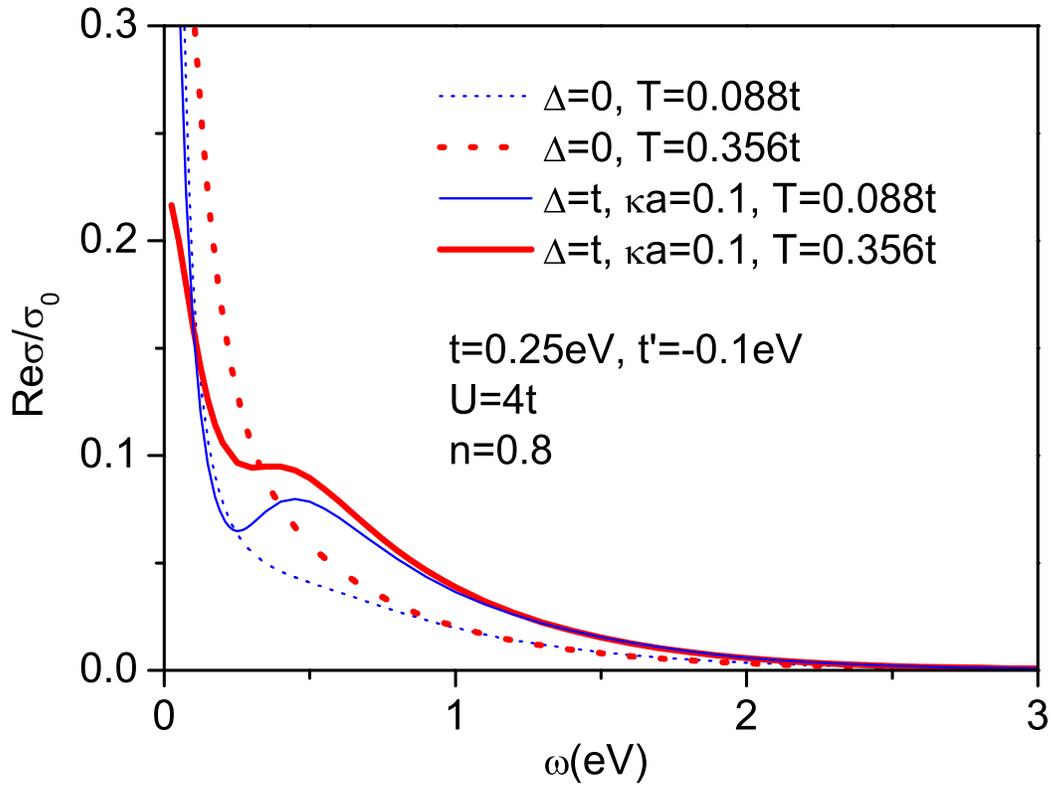}
\caption{(Color online) Real part of optical conductivity for correlated 
metal ($U=4t$, $t'=-0.4t$, $t=0.25$ eV) in 
DMFT+ $\Sigma_{\bf p}$ approximation for two different temperatures:
$T=0.088t$ and $T=0.356t$. Pseudogap amplitude $\Delta=t$, correlation
length $\xi=10a$, filling factor $n=0.8$ electrons per atom.}
\label{DMFT_S_T}
\end{figure}

\begin{figure}
\includegraphics[clip=true,width=1.0\textwidth]{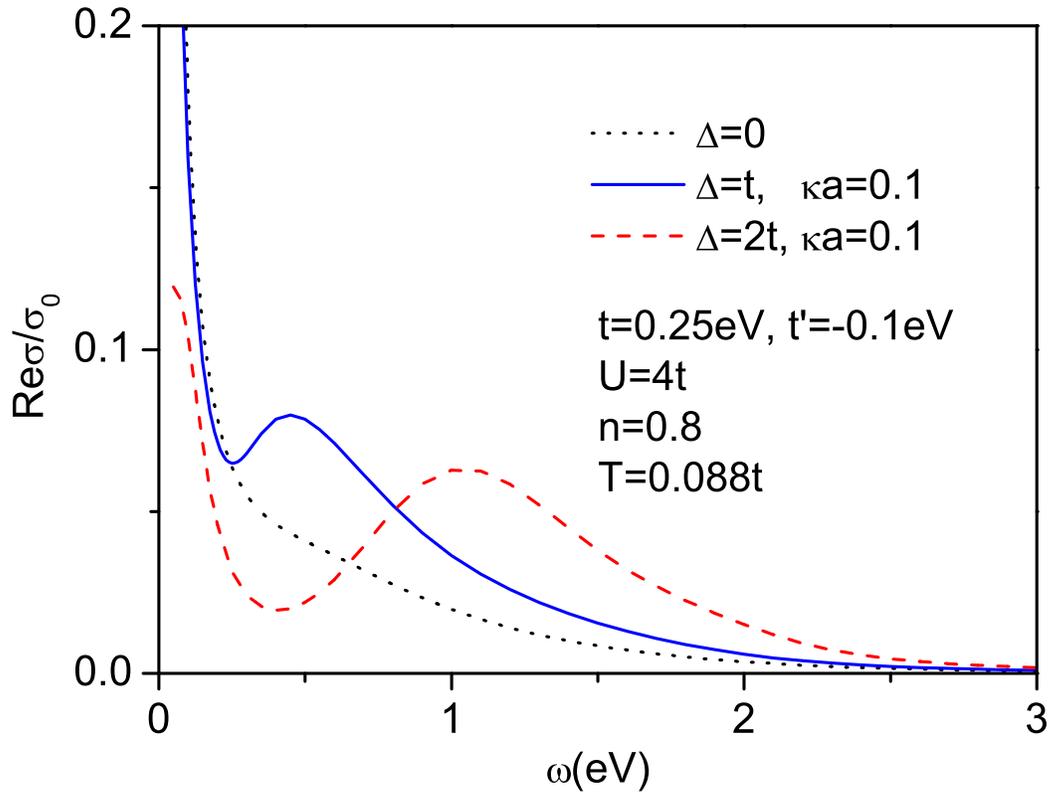}
\caption{(Color online) Real part of optical conductivity for correlated 
metal ($U=4t$, $t'=-0.4t$, $t=0.25$ eV) in 
DMFT+$\Sigma_{\bf p}$ approximation --- $\Delta$ dependence.  
Parameters are the same as in Fig. \ref{DMFT_S_T}, but data are for different 
values of $\Delta=0$, $\Delta=t$, $\Delta=2t$, and temperature 
$T=0.088t$. } 
\label{DMFT_S_D} 
\end{figure}

\begin{figure}
\includegraphics[clip=true,width=1.0\textwidth]{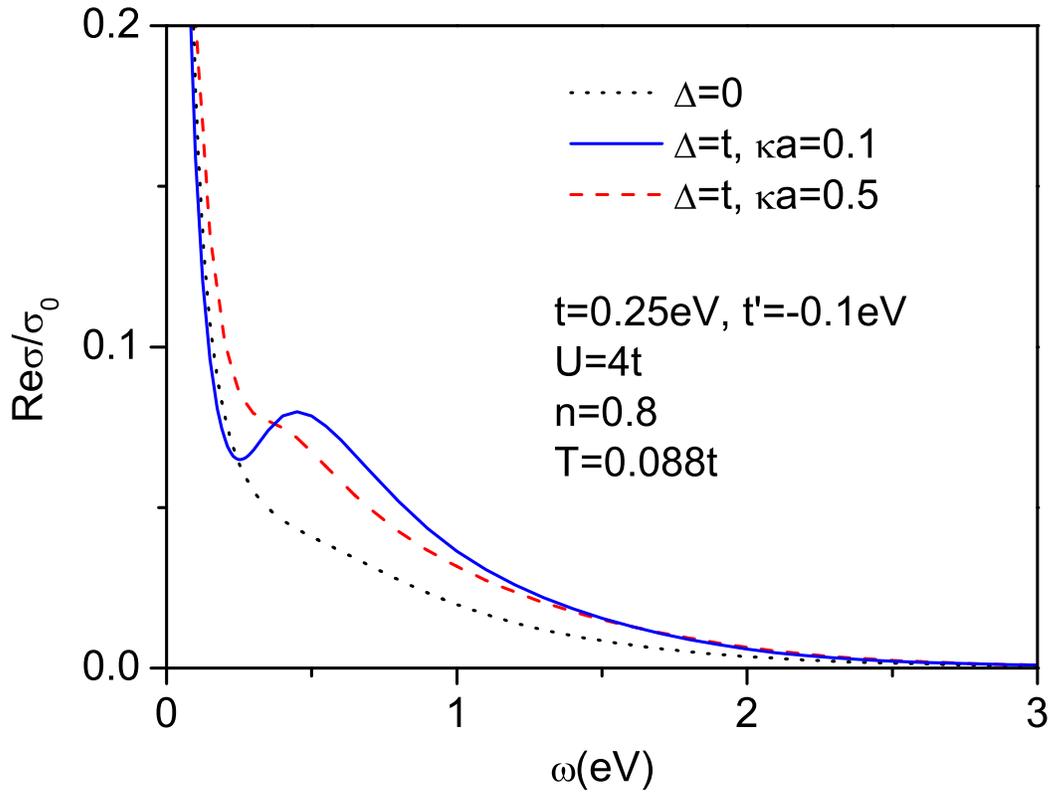}
\caption{(Color online) Real part of optical conductivity for correlated 
metal ($U=4t$, $t'=-0.4t$, $t=0.25$ eV) in 
DMFT+$\Sigma_{\bf p}$ approximation --- dependence on correlation length.
Parameters are the same as in Fig. \ref{DMFT_S_T}, but data are for 
different values of inverse correlation length
$\kappa=\xi^{-1}$: $\kappa a=0.1$ and $\kappa a=0.5$, and 
temperature $T=0.088t$.} 
\label{DMFT_S_kap} 
\end{figure}

\begin{figure}
\includegraphics[clip=true,width=0.95\textwidth]{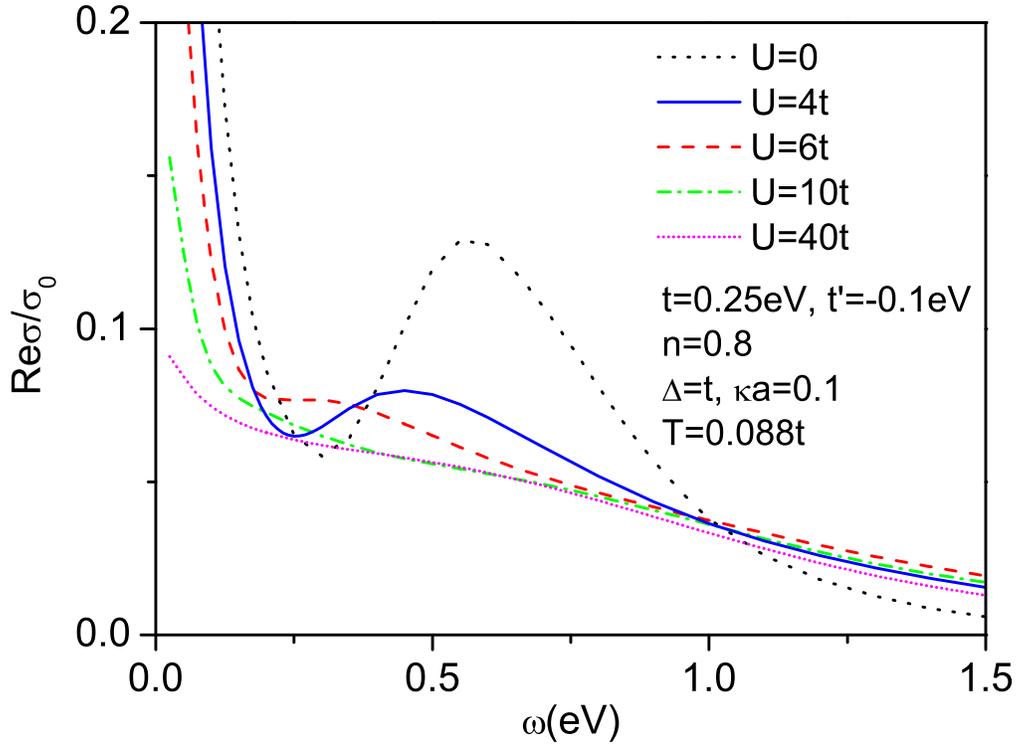}
\caption{(Color online) Real part of optical conductivity for correlated 
metal in DMFT+$\Sigma_{\bf p}$ approximation --- $U$ dependence.
Parameters are the same as in Fig. \ref{DMFT_S_T}, but data are for 
different values $U$: $U=0$, $U=4t$, $U=6t$, $U=10t$ and
$U=40t$. Temperature $T=0.088t$.}
\label{DMFT_S_U}
\end{figure}

In Fig. \ref{DMFT_S_T} we show our DMFT+ $\Sigma_{\bf p}$ results for the 
real part of optical conductivity for correlated metal ($U=4t$) for two
values of temperature, compared with similar data without pseudogap
fluctuations (pure DMFT). We clearly observe formation of typical 
pseudogap (absorption) anomaly on the ``shoulder'' of Drude--like peak, 
which is partially ``filled'' with the growth 
of temperature. This behavior is quite similar to ``mid-infrared feature'' 
that is observed in optical conductivity of cuprate superconductors 
\cite{Bas,Timu}. In Fig. \ref{DMFT_S_D} we show the behavior of Re$\sigma(\omega)$
for different values of the pseudogap
amplitude $\Delta$. We see that pseudogap anomaly naturally grows with the
growth of $\Delta$. Fig. \ref{DMFT_S_kap} illustrates the dependence of 
Re$\sigma(\omega)$ on correlations length of pseudogap (AFM,
SDW) fluctuations. Again we observe the natural behavior --- pseudogap
anomaly is ``filled'' for shorter correlation lengths, i.e. as fluctuations
become more short--ranged. At last, in Fig. \ref{DMFT_S_U} we demonstrate
dependence of pseudogap anomaly in optical conductivity on correlation 
strength, i.e. on Hubbard interaction $U$. It is seen that the frequency 
range, where pseudogap anomaly is observed becomes narrower as correlation
strength grows. This correlates with general narrowing of the pseudogap 
anomaly and spectral densities with the growth of correlations, observed in 
our previous work \cite{PRB05,FNT06}. For large values of $U$ pseudogap 
anomaly is practically suppressed. This is the main qualitative difference
of the results of the present approach, compared to our earlier work 
\cite{SS02} on optical conductivity in the pseudogap state. Comparing 
the data of present work for $U=0$ with similar data of Ref.\cite{SS02}, it 
should be noted, that in this earlier work we have performed calculations of 
dynamic conductivity only for $T=0$ and used simplified expressions, neglecting 
$RR,AA$--loops contributions to conductivity, as well as small frequency 
expansion \cite{VW}, just to speed up calculations. These simplifications lead 
to some quantitative differences with the results of present work, where all 
calculations are done exactly using the general expression (\ref{cond_final}), 
though qualitatively the frequency behavior of conductivity is the same.

\subsubsection{Doped Mott insulator}

Now we shall discuss our results for the case of doped Mott insulator with
$U\gg W$. This case has no direct relevance to copper oxides, but is 
interesting from the general point of view and we present some of our results.

\begin{figure}
\includegraphics[clip=true,width=1.0\textwidth]{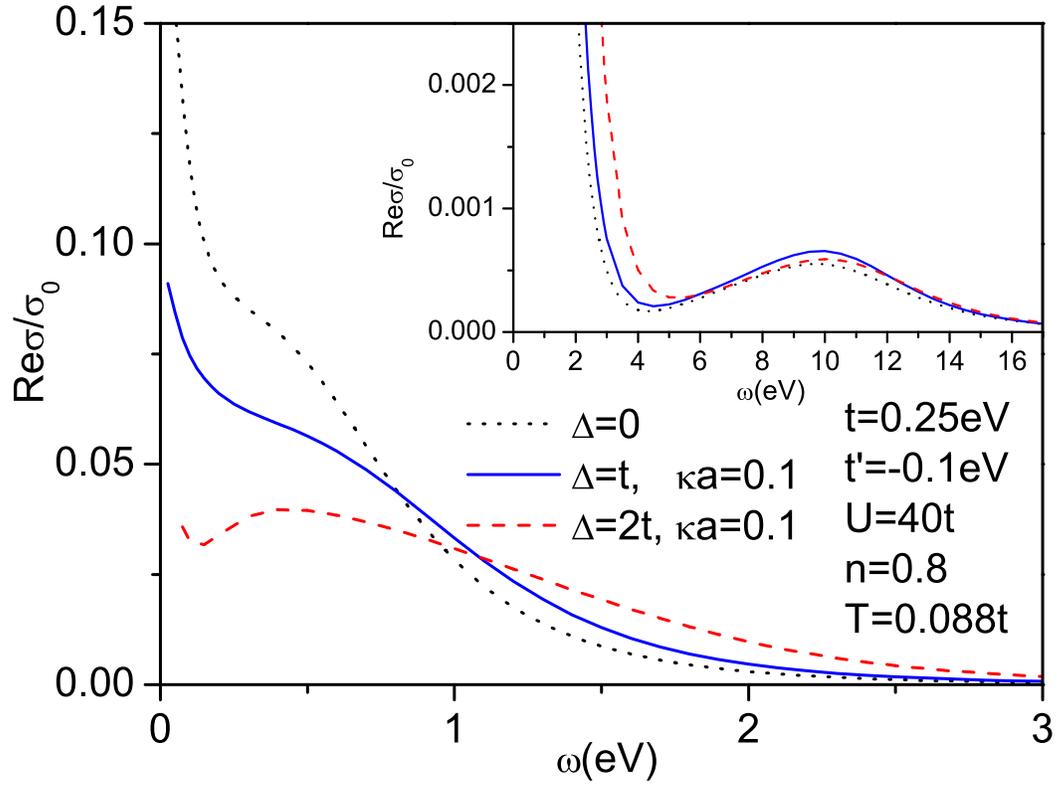}
\caption{(Color online) Real part of optical conductivity for doped Mott 
insulator ($U=40t$, $t'=-0.4t$, $t=0.25$ eV) in DMFT+$\Sigma_{\bf p}$ 
approximation for different values of $\Delta=0$, $\Delta=t$, $\Delta=2t$, 
and temperature $T=0.088t$. Correlation length $\xi=10 a$, 
filling factor $n=0.8$.
Insert: conductivity in a wide frequency interval, including transitions to
the upper Hubbard band.} 
\label{DMFT_S_D_i} 
\end{figure}

\begin{figure}
\includegraphics[clip=true,width=1.0\textwidth]{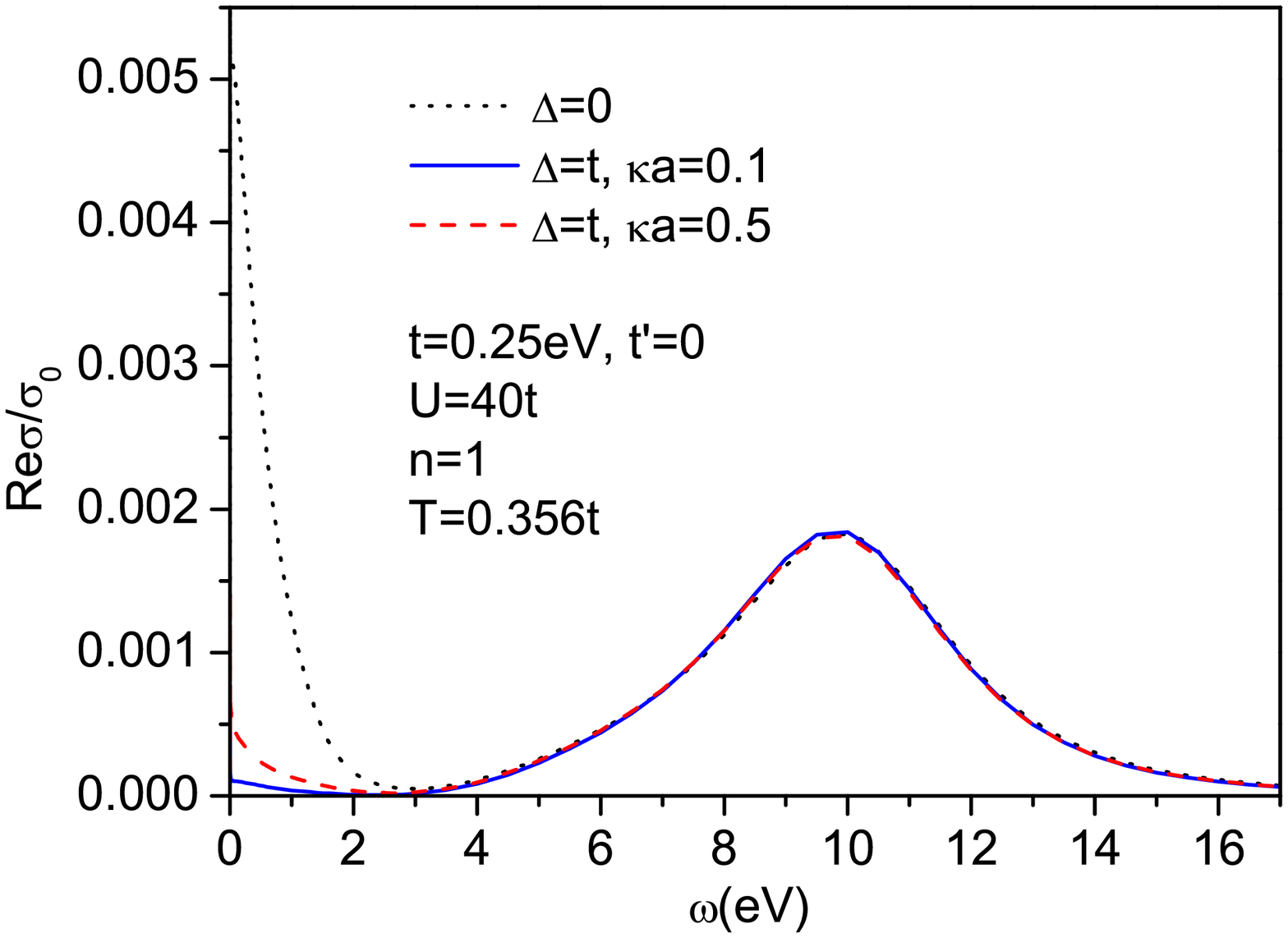}
\caption{(Color online) Real part of optical conductivity for doped Mott 
insulator ($U=40t$, $t=0.25$ eV$, t'=0$) in DMFT+$\Sigma_{\bf p}$ approximation for different 
values of inverse correlation length $\kappa=\xi^{-1}$: $\kappa a=0.1$ and 
$\kappa a=0.5$, temperature $T=0.356t$ and filling $n=1$.}  

\label{DMFT_S_kap_i} 
\end{figure}

The real part of optical conductivity for the case of
$U=40t$ is shown in Figs. \ref{DMFT_S_D_i},\ref{DMFT_S_kap_i}.

In Fig. \ref{DMFT_S_D_i} we show Re$\sigma(\omega)$ for doped Mott insulator 
in DMFT+$\Sigma_{\bf p}$ approach several values of pseudogap amplitude
$\Delta$. Obviously enough, pseudogap fluctuations lead to significant
changes of optical conductivity only for relatively small frequencies of
the order of $\Delta$, while for high frequencies (e.g. of the order of $U$,
where the upper Hubbard band contributes) we do not observe pseudogap 
effects (see insert in Fig. \ref{DMFT_S_D_i}). For small frequencies we
observe pseudogap suppression of Drude--like peak, with only a shallow anomaly
for $\omega\sim\Delta$, which just disappears for smaller values of $\Delta$
or shorter correlation lengths. 

In Fig. \ref{DMFT_S_kap_i} we show the similar data for
the special case of $t'=0$ and $n=1$, i.e. at half--filling (Mott insulator)
for different values of inverse correlation length $\kappa=\xi^{-1}$. 
Conductivity at small frequencies is determined only by thermal excitations 
and pseudogap fluctuations suppress it significantly. Shorter correlation
lengths obviously lead to larger values of conductivity at small frequencies.
Transitions to upper Hubbard band are not affected by these fluctuations at 
all.

\section{Conclusion}
\label{concl}

The present work is the direct continuation of our previous work 
\cite{JTL05,PRB05,FNT06}, where we have proposed a generalized 
DMFT+$\Sigma_{\bf p}$ approach, which is meant to take into account the 
important effects of non--local correlations (in principle of any type) in 
addition to the (essentially exact) treatment of local dynamical correlations 
by DMFT. Here we used a generalized DMFT+$\Sigma_{\bf p}$ approach to 
calculate dynamic (optical) conductivity of two-dimensional Hubbard model 
with pseudogap fluctuations. Our results demonstrate, that pseudogap 
anomalies observed in optical conductivity of copper oxides can, in 
principle, be explained by this model.
Main advantage in comparison to the previous work \cite{SS02} is our
ability now to study the role of strong electronic correlations, which are
decisive in the formation of electronic structure of systems like copper
oxides. In fact, we have demonstrated an important suppression of pseudogap
anomaly in optical conductivity with the growth of correlation strength.

As we already noted in Ref. \cite{PRB05} qualitatively similar results on 
pseudogap formation in single--particle characteristics for the 2d Hubbard 
model were also obtained within cluster extensions of DMFT 
\cite{TMrmp,KSPB}. However, these methods have generic restrictions 
concerning the size of the cluster, and up to now were not widely applied to 
calculations of two--particle properties, such as general response 
functions, and, in particular, to calculations of dynamic (optical)
conductivity.

Our approach is free of these limitations, though for the price of
introduction of additional (semi) phenomenological parameters (correlation 
length $\xi$, and pseudogap amplitude $\Delta$). It is mauch less time 
consuming, thus its advantage for calculations of two--particle response 
functions is obvious. 
It also opens the possibility of systematic comparison of different 
types of non-local fluctuations and their effects on electronic properties,
providing more intuitive way to analyze experiments or theoretical data 
obtained within more advanced schemes. 
Note, again, that in principle both $\xi$ and $\Delta$ can be calculated from 
the original model\cite{PRB05}.
Our scheme works for any Coulomb interaction strength $U$,
pseudogap strength $\Delta$, correlation length $\xi$, filling $n$
and bare electron dispersion $\varepsilon({\bf k})$. 

The present formalism can be easily generalized in the framework of 
our recently proposed LDA+DMFT+$\Sigma_{\bf p}$ approach, which will allow to 
perform calculations of pseudogap anomalies of optical conductivity for
realistic models. It can also be easily generalized to orbital degrees of 
freedom, phonons, impurities, etc.

\section{Acknowledgements}

We are grateful to Th. Pruschke for providing us with his effective NRG code.
This work was supported in part by RFBR grants 05-02-16301 (MS,EK,IN),
05-02-17244 (IN), 06-02-90537 (IN), by the joint UrO-SO project (EK,IN), and programs of 
the Presidium of the Russian Academy of Sciences (RAS) ``Quantum macrophysics''
and of the Division of Physical Sciences of the RAS ``Strongly correlated
electrons in semiconductors, metals, superconductors and magnetic
materials''. I.N. acknowledges support from the Dynasty Foundation and 
International Center for Fundamental Physics in Moscow program for young
scientists and also from the grant of President of Russian Federation for
young PhD MK-2118.2005.02.



\appendix

\section{Ward identities}
\label{A}

In this Appendix we present derivation of Ward identities used in the main text.
Let us start with the general expression for variation of electron 
self--energy due to an arbitrary variation of the complete Green's function,
which is valid for any interacting Fermi system \cite{Mig}:
\begin{equation}
\Delta\Sigma_p=\sum_{p'}U_{pp'}(q)\Delta G_{p'},
\label{WMig}
\end{equation}
where $U_{pp'}(q)$ is an irreducible vertex in particle--hole channel, and we
use 4-dimensional notations $p=(i\varepsilon,{\bf p})$, $q=(i\omega,{\bf q})$ 
etc. In the following we take:
\begin{equation}
\Delta\Sigma_p=\Sigma_+-\Sigma_-\equiv \Sigma(i\varepsilon_+,{\bf p}_+)-
\Sigma(i\varepsilon_-,{\bf p}_-)
\label{DSigma}
\end{equation}
and (in the same notations): 
\begin{equation}
\Delta G_p=G_+-G_-=(G_+G_-)_p(\Delta\Sigma_p-\Delta(G_0^{-1})_p),
\label{DGr}
\end{equation}
where $\Delta(G_0^{-1})_p=G^{-1}_{0+}-G^{-1}_{0-}$, and the last expression was
obtained using the standard Dyson equation.

Note the similarity of Eq. (\ref{WMig}) to the Ward identity for non-interacting
electrons in the impure system, derived in Ref. \cite{VW}.

Now substituting the last expression in (\ref{DGr}) we get:
\begin{equation}
\Delta\Sigma_{p}=\sum_{p'}U_{pp'}(q)(G_+G_-)_{p'}(\Delta\Sigma_{p'}
-\Delta(G^{-1}_{0})_{p'}).
\label{DGreen}
\end{equation}
Iterating this equation we obtain:
\begin{eqnarray}
\Delta\Sigma_p=\sum_{p'}U_{pp'}(G_+G_-)_{p'}(-\Delta(G^{-1}_{0})_{p'})+\nonumber\\
+\sum_{p''p'}U_{pp''}(G_+G_-)_{p''}U_{p''p'}(G_+G_-)_{p'}
(-\Delta(G^{-1}_0)_{p'})+...~~.
\label{itW}
\end{eqnarray}
Multiplying both sides of (\ref{itW}) by $(G_+G_-)_{p}$ and adding 
\begin{equation}
\sum_{p'}(G_+G_-)_{p}\delta_{pp'}(-\Delta(G^{-1}_0)_{p'})=
(G_+G_-)_{p}(-\Delta(G^{-1}_0)_p)
\nonumber
\end{equation}
we have:
\begin{eqnarray}
(G_+G_-)_{p}(\Delta\Sigma_p-\Delta(G^{-1}_0)_{p})=\sum_{p'}[(G_+G_-)_p
\delta_{pp'}+(G_+G_-)_{p}U_{pp'}(G_+G_-)_{p'}+\nonumber\\
+(G_+G_-)_p\sum_{p''}U_{pp''}(G_+G_-)_{p''}U_{p''p'}(G_+G_-)_{p'}+...]
(-\Delta(G^{-1}_0))=\nonumber\\
=\sum_p\Phi_{pp'}(q)(-\Delta(G^{-1}_0)_{p'}),
\label{itrW}
\end{eqnarray}
where $\Phi_{pp'}(q)$ is the complete two -- particle Green's function determined
by the following Bethe -- Salpeter equation \cite{Mig}:
\begin{equation}
\Phi_{pp'}(q)=(G_+G_-)_p\delta_{pp'}+(G_+G_-)_p\sum_{p'}U_{pp'}\Phi_{pp'}(q).
\label{BetSal}
\end{equation}
Finally we obtain:
\begin{equation}
\Delta G_{p}=\sum_{p'}\Phi_{pp'}(q)(-\Delta(G^{-1}_0)_{p'}),
\label{Wid}
\end{equation}
which the general form of our Ward identity.

Summing both sides of (\ref{Wid}) over ${\bf p}$ and taking ${\bf q}=0$ we
obtain the identity (\ref{Ward1}) used above. Similarly, taking the ``bare''
Green's function (\ref{G_DMFT}) we obtain (\ref{Ward2}).

\pagestyle{empty}

\newpage

\begin{thebibliography}{99}
\vfill

\bibitem{Tim}T. Timusk, B. Statt, Rep. Progr. Phys, {\bf 62}, 61 (1999).

\bibitem{MS}M. V. Sadovskii, Usp. Fiz. Nauk {\bf 171}, 539 (2001)
[Physics--Uspekhi {\bf 44}, 515 (2001)].

\bibitem{Sch}J. Schmalian, D. Pines, B.Stojkovi\v c, 
Phys. Rev. Lett. {\bf 80}, 3839 (1998); Phys. Rev. B {\bf 60}, 667 (1999).

\bibitem{KS}E. Z. Kuchinskii, M. V. Sadovskii, Zh. Eksp. Teor. Fiz. {\bf 115},
1765 (1999) [(JETP {\bf 88}, 347 (1999)].

\bibitem{MetzVoll89}  W. Metzner and D. Vollhardt, Phys. Rev. Lett.
{\bf 62}, 324 (1989).

\bibitem{vollha93}  D.~Vollhardt, in {\em Correlated Electron Systems},
edited by V.~J. Emery, World Scientific, Singapore, 1993, p.~57.

\bibitem{pruschke}  Th. Pruschke, M. Jarrell, and J. K. Freericks, Adv.
in Phys. {\bf 44}, 187 (1995).

\bibitem{georges96}  A. Georges, G. Kotliar, W. Krauth, and M. J. Rozenberg,
Rev. Mod. Phys. {\bf 68}, 13 (1996).

\bibitem{PT} G.\ Kotliar and D.\ Vollhardt, Physics Today \textbf{57},
No.\ 3 (March), 53 (2004).

\bibitem{Si96}Q. Si and J.L. Smith, Phys. Rev. Lett. {\bf 77}, 3391 (1996).

\bibitem{Haule} EDMFT approach to pseudogap formation can be found in  
K. Haule, A. Rosch, J. Kroha, and P. W\"olfle, Phys. Rev. Lett. \textbf{89}, 236402 (2002);
K. Haule, A. Rosch, J. Kroha, and P. W\"olfle, Phys. Rev. B \textbf{68}, 155119 (2003).

\bibitem{TMrmp} Th.\ Maier, M.\ Jarrell, Th.\ Pruschke and M.\ Hettler,
Rev.\ Mod.\ Phys.\ Rev. Mod. Phys. {\bf 77}, 1027 (2005).

\bibitem{KSPB}G. Kotliar, S.Y. Savrasov, G. Palsson, G. Biroli,
Phys. Rev. Lett. {\bf 87}, 186401 (2001); M. Capone, M. Civelli, 
S.S. Kancharla, C. Castellani, and G. Kotliar, 
Phys.\ Rev.\ B {\bf 69}, 195105 (2004).

\bibitem{JTL05}E.Z.Kuchinskii, I.A.Nekrasov, M.V.Sadovskii. Pis'ma Zh. Eksp.
Teor. Fiz. {\bf 82}, 217 (2005) 
[JETP Lett. {\bf 82}, 198 (2005)].

\bibitem{PRB05}M.V. Sadovskii, I.A. Nekrasov, E.Z. Kuchinskii, Th. Prushke,
V.I. Anisimov. Phys. Rev. B {\bf 72}, 155105 (2005).

\bibitem{FNT06}E.Z. Kuchinskii, I.A. Nekrasov, M.V. Sadovskii. Fizika Nizkikh
Temperatur {\bf 32}, 528 (2006)
[Low Temp. Phys. {\bf 32}, 398 (2006)].

\bibitem{cm06}E.Z. Kuchinskii, I.A. Nekrasov, Z.V. Pchelkina, M.V. Sadovskii.
ArXiv:\ cond-mat/0606651.

\bibitem{VW}D.Vollhardt, P.W\"olfle. Phys. Rev. B {\bf 22}, 4666 (1980).

\bibitem{Diagr}M.V. Sadovskii. Diagrammatics. World Scientific, Singapore 2006.

\bibitem{Janis}V. Jani\v s, J. Koloren\v c, V. \v Spi\v cka. Eur. J. Phys. B
{\bf 35}, 77 (2003).

\bibitem{MS79}M. V. Sadovskii, Zh. Eksp. Teor. Fiz. {\bf 77}, 2070(1979)
[Sov.Phys.--JETP {\bf 50}, 989 (1979)].

\bibitem{ST91}M.V. Sadovskii, A.A. Timofeev.  J. Moscow Phys. Soc. 
{\bf 1}, 391 (1991).

\bibitem{SS02}M.V. Sadovskii, N.A. Strigina. Zh. Eksp. Teor. Phys. {\bf 122},
610 (2002) [JETP {\bf 95}, 526 (2002)].

\bibitem{S74} M.V.Sadovskii. Zh. Eksp. Teor. Fiz. {\bf 66}, 1720 (1974))
[Sov. Phys. -- JETP {\bf 39}, 845 (1974)]. 

\bibitem{NRG} K.G.~Wilson,  Rev.\ Mod.\ Phys. {\bf 47}, 773 (1975);\ 
H.R.~Krishna-murthy, J.W.~Wilkins, and K.G.~Wilson, Phys.\ Rev.\ B {\bf 21}, 
1003 (1980); {\it ibid.} {\bf 21}, 1044 (1980);\ A.C.\ Hewson, {\em The 
Kondo Problem to Heavy Fermions} (Cambridge University Press, 1993).  

\bibitem{BPH}R.~Bulla,  A.C.~Hewson and Th.~Pruschke,
J.\ Phys.\ -- Condens.\ Matter {\bf 10}, 8365(1998); 
R.~Bulla, Phys.\ Rev.\ Lett. {\bf 83}, 136 (1999).

\bibitem{Bas}D.N. Basov, T. Timusk. Rev. Mod. Phys. {\bf 77}, 721 (2005).

\bibitem{Timu}J. Hwang, T. Timusk, G.D. Gu. ArXiv:\ cond-mat/0607653.

\bibitem{Mig}A.B. Migdal. Theory of Finite Fermi Systems and Applications to
Atomic Nuclei. Interscience Publishers. NY 1967.



\end{thebibliography}
\end{document}